\documentclass[reprint,aps]{revtex4}
\usepackage{amsmath}
\usepackage{graphicx}
\usepackage{dcolumn}
\usepackage{bm}

\usepackage[colorlinks=true,allcolors=blue]{hyperref}
\usepackage{amssymb}
\usepackage{dsfont}
\usepackage{url}
\usepackage{caption,subcaption}

\begin{document}

\title{Transient fluid dynamics with general matching conditions: a first study from the method of moments}
\author{Gabriel~S.~Rocha}
\author{Gabriel S. Denicol}
\affiliation{Instituto de F\'{\i}sica, Universidade Federal Fluminense, Niter\'{o}i, Rio de Janeiro,
Brazil}

\begin{abstract}
Recent works have revealed that matching conditions play a major role on general consistency properties of relativistic fluid dynamics such as causality, stability and wellposedness of the equations of motion. In this paper we derive transient fluid dynamics from kinetic theory, using the method of moments as proposed by Israel and Stewart, without imposing an specific matching condition. We then investigate how the equations of motion and their corresponding transport coefficients are affected by the choice of matching condition. 
\end{abstract}

\maketitle

\email{gabrielsr@id.uff.br}

\email{gsdenicol@id.uff.br}

\section{Introduction}
Relativistic fluid dynamics is an effective theory derived to describe the long-distance, long-time dynamics of a given microscopic theory. It is applied in several fields of physics, from the description of neutron star mergers \cite{rezzolla2013relativistic,Takami:2014tva} to the hot and dense nuclear matter produced in ultra-relativistic heavy ion collisions \cite{yagi2005quark,Heinz:2013th,Gale:2013da}. Nevertheless, the theoretical foundations of relativistic \textit{dissipative} fluid dynamics remain open, still being topic of intense investigation \cite{Florkowski:2017olj, romatschke2019relativistic}.

Relativistic generalizations of Navier-Stokes theory display physical and mathematical pathologies which practically prevents its use in general applications. The main issue is that Navier-Stokes theory is acausal and displays intrinsic instabilities \cite{hiscock:85generic} when perturbed around a global equilibrium state -- an illness that cannot be corrected by adjusting matching conditions or transport coefficients. An early attempt to derive a linearly causal and stable fluid-dynamical theory was put forward by Israel and Stewart in the 1970's \cite{israel1979annals,israel1979jm}. The main feature of Israel-Stewart theory is that, instead of imposing constitutive relations relating the dissipative currents with derivatives of the velocity field, they promoted the non-equilibrium fields to independent dynamical variables for which they derived relaxation equations. Then, the requirements of linear causality and stability yield constrains on the relaxation times that are allowed in the formulation \cite{hiscock1983stability,denicol2008stability,brito2020linear,biswas2020causality}. Even so, in Israel-Stewart theories, existence and uniqueness of solutions are not guaranteed in general curved spaces \cite{bemfica:18causality}. 

Recently, Bemfica, Disconzi and Noronha proposed a novel theory of fluid dynamics which is not derived following the procedure outlined by Israel and Stewart \cite{bemfica:20general,bemfica:19nonlinear,bemfica:18causality}. Other aspects about this formulation were also investigated in Refs.\ \cite{kovtun:19first,hoult2020stable}. In this approach, fluid dynamics is constructed from a generalized gradient expansion, which includes both time-like and space-like gradients. This is in contrast to the traditional approach, used to derive Navier-Stokes theory, which consists of an expansion only in space-like gradients. It was then demonstrated that a first order theory in such generalized gradient expansion can lead to a causal and linearly stable theory as long as appropriate matching conditions are imposed.

Matching conditions are constraints required to define the local equilibrium state of a fluid in the presence of dissipation, i.e., they define the temperature, chemical potential, and velocity of a viscous fluid. These conditions are an essential feature of relativistic dissipative fluid dynamics, since such theories are always constructed in terms of an expansion around a \textit{fictitious} local equilibrium state. The most well-known matching conditions are those due to Landau and Lifshitz \cite{landau:59fluid}, often employed in simulations of ultra-relativistic heavy ion collisions \cite{Shen:2014vra,NunesdaSilva:2020bfs,Schenke:2010rr}, and those due to Eckart \cite{eckart:40fluid}, more convenient for Astrophysics applications \cite{chabanov:21-general}. However, the novel fluid-dynamical theory derived by Bemfica, Disconzi and Noronha was shown to be acausal and linearly unstable exactly for those matching conditions. This motivated the search and understanding of new matching conditions and their effect on the dynamics of relativistic fluids.    

In particular, the traditional Israel-Stewart equations must also display a dependence on matching conditions. This dependence has been initially explored using a phenomenological derivation of fluid dynamics from the second law of thermodynamics \cite{noronha2021transient}. Given this scenario, it is also relevant to assess how the derivation of transient fluid dynamics from kinetic theory, using the method of moments as proposed by Israel and Stewart, is affected by the different choices of matching. Extending this derivation to be applicable to general matching conditions and investigating their effects is the main goal of this paper.

The present text is organized as follows. In Sec.~\ref{sec:1-fluid} we give an overview of the fluid-dynamical variables when general matching conditions are employed. In section \ref{sec:2-BEq-fluid-vars}, we relate the general fluid-dynamical variables to the single particle distribution function. In Sec.~\ref{sec:3-Moment-eqs} we proceed to derive the system of coupled equations for the moments of the single-particle distribution function, which is later, in Sec.~\ref{sec:4-truncation-19} truncated using what we refer to as the 19 moments approximation -- a generalization of the 14-moment approximation for general matching conditions. Next, in Sec.~\ref{sec:5-collisionterm} we use the relaxation time approximation proposed in Ref.\ \cite{rocha:21} to simplify the collision integrals that appear in the derivation of fluid dynamics from the Boltzmann equation. The final equations of motion to the fluid-dynamic variables are portrayed in \ref{sec:6-EoMs-m=0} in the massless limit. Then, in Sec.~\ref{sec:7-exotic-eckart} we proceed to give a concrete example of the results for a restricted class of matching conditions for which one has vanishing particle density and particle diffusion, which we label \textit{Exotic Eckart conditions}. Section \ref{sec:conclusion} concludes the text. 

\textbf{Notation and conventions}:
We use $(+ - - -)$ as the metric signature; natural units, so that $\hbar = c = k_{B} = 1$.

\section{Fluid-dynamical variables}
\label{sec:1-fluid}

The main fluid-dynamical equations are the continuity equations related to the conservation of net-charge, energy, and momentum,
\begin{equation}
\begin{aligned}
\label{eq:consv-eqns}
    \partial_{\mu}N^{\mu} = 0,\\
    \partial_{\mu}T^{\mu \nu} = 0,
\end{aligned}
\end{equation}
where $N^{\mu}$ is the net-charge 4-current and $T^{\mu \nu}$ is the energy-momentum tensor. 

The traditional fluid-dynamical variables are defined from the tensor decomposition of $N^{\mu}$ and $T^{\mu \nu}$ in terms of a time-like 4-vector, $u^{\mu}$ -- the fluid 4-velocity. This 4-vector is assumed to be unitary, $u_{\mu}u^{\mu}=1$, and shall be formally defined later using matching conditions. An irreducible tensor decomposition of these fields read    
\begin{equation}
\begin{aligned}
\label{eq:decompos-numu-tmunu}
   N^{\mu} &= n u^{\mu} + \nu^{\mu},  \\
    T^{\mu \nu} &= \varepsilon u^{\mu} u^{\nu} - P \Delta^{\mu \nu} + h^{\mu} u^{\nu} + h^{\nu} u^{\mu} + \pi^{\mu \nu},
\end{aligned}
\end{equation}
where $n$, $\varepsilon$, and $P$ are the net-charge density, energy density, and isotropic pressure in the local rest frame of the fluid ($u^{\mu}=(1,0,0,0)$), respectively. Furthermore, we introduced the net-charge and energy diffusion currents, $\nu^{\mu}$ and $h^{\mu}$, respectively, and the shear-stress tensor, $\pi^{\mu\nu}$. For the sake of convenience, we also defined the projection operator onto the 3-space orthogonal to $u^\mu$, $\Delta^{\mu\nu} \equiv g^{\mu\nu} - u^\mu u^\nu$. These fields can be identified in terms of the following projections of the conserved currents,
\begin{equation}
\begin{aligned}
\label{eq:definitions}
    n &\equiv u_{\mu}N^{\mu} , \, \, \varepsilon \equiv u_{\mu}u_{\nu}T^{\mu\nu}, \, \, P \equiv -\frac{1}{3}\Delta_{\mu\nu}T^{\mu\nu},  \\
    \nu^{\mu} &\equiv \Delta^{\mu}_{\nu} N^{\nu}, \, \, h^{\mu} \equiv \Delta^{\mu}_{\nu} u_{\lambda} T^{\nu\lambda}, \, \, \pi^{\mu\nu} \equiv \Delta^{\mu\nu}_{\alpha\beta} T^{\alpha\beta} . \\
\end{aligned}
\end{equation}
Above, we introduced the double, traceless, and symmetric projection operator $\Delta^{\mu \nu \alpha \beta} = \frac{1}{2}\left( \Delta^{\mu \alpha } \Delta^{\nu \beta} + \Delta^{\nu \alpha } \Delta^{\mu \beta} \right) - \frac{1}{3} \Delta^{\mu \nu} \Delta^{\alpha \beta}$.

Next, we define a fictitious equilibrium state with  inverse temperature, $\beta \equiv 1/T$, and thermal potential, $\alpha \equiv \mu/T$, where $\mu$ is the chemical potential and $T$ is the temperature. We then decompose the net-charge density, energy density and isotropic pressure as
\begin{equation}
\begin{aligned}
\label{eq:definitions2}
    n &\equiv n_0(\alpha,\beta) + \delta n, \\
    \varepsilon &\equiv \varepsilon_0(\alpha,\beta) + \delta \varepsilon, \\ 
    P &\equiv P_0(\alpha,\beta) + \Pi,
\end{aligned}
\end{equation}
where $n_0$, $\varepsilon_0$ and $P_0$ are the equilibrium net-charge density, energy density, and pressure, respectively, and are determined from $\alpha$ and $\beta$ using an equation of state. In this case, $\delta n$ is a non-equilibrium correction to the net-charge density, $\delta \varepsilon$ is a non-equilibrium correction to the energy density, and $\Pi$ is the bulk viscous pressure. Naturally, at this point these dissipative quantities are not specified, since it was not determined how the separation of the net-charge and energy density into an equilibrium part and a non-equilibrium part is implemented. This will be done by introducing additional constrains or matching conditions (which will complement those that will be introduced to define the fluid 4-velocity). Once this task is performed, the temperature and chemical potential are obtained by inverting the thermodynamic functions, $n_0(\alpha,\beta)$ and $\varepsilon_0(\alpha,\beta)$. Once these quantities are known, one can calculate the thermodynamic pressure and obtain $\Pi$.

The most well known and used matching conditions are those proposed by Landau \cite{landau:59fluid} and Eckart \cite{eckart:40fluid}. In the Landau picture, the 4-velocity is defined as an Eigenvector of $T^{\mu}_{\ \nu}$,
\begin{equation}
\label{eq:Landau}
T^{\mu}_{\ \nu} u^{\nu} \equiv \varepsilon u^\mu \Longleftrightarrow h^\mu \equiv 0, 
\end{equation}
which effectively eliminates any energy diffusion. While in the Eckart picture, the same quantity is defined from the direction of the net-charge 4-current,
\begin{equation}
\label{eq:Eckart}
N^{\mu} \equiv n u^{\mu} \Longleftrightarrow \nu^\mu \equiv 0, 
\end{equation}
which effectively eliminates any net-charge diffusion. In both the Eckart and Landau pictures, the temperature and chemical potential are defined in the same way, by simply eliminating any non-equilibrium corrections to $n$ and $\varepsilon$,  
\begin{equation}
\begin{aligned}
\label{eq:decompos-numu-tmunu2}
   n &\equiv n_0(\alpha,\beta) 
   \Longleftrightarrow \delta n \equiv 0, \\
   \varepsilon &\equiv \varepsilon_0(\alpha,\beta) 
   \Longleftrightarrow \delta \varepsilon \equiv 0 .
\end{aligned}
\end{equation}
We note that one can also use other types of matching conditions, even though they become rather nontrivial to define. In particular, this task becomes rather complicated if the matching conditions use fields which are not among the conserved currents. Nevertheless, this task can always be accomplished in kinetic theory and will be explained in the following section. 

In summary, for an arbitrary matching condition the conserved currents are decomposed as
\begin{equation}
\begin{aligned}
\label{eq:decompos-numu-tmunu3}
   N^{\mu} &= (n_{0} + \delta n) u^{\mu} + \nu^{\mu}  \\
    T^{\mu \nu} &= (\varepsilon_{0} + \delta \varepsilon )u^{\mu} u^{\nu} - (P_{0} + \Pi) \Delta^{\mu \nu} + h^{\mu} u^{\nu} + h^{\nu} u^{\mu} + \pi^{\mu \nu}.
\end{aligned}
\end{equation}
This tensor decomposition describes the conserved currents using a total of 19 degrees of freedom -- 5 more than the 14 independent components of $N^{\mu}$ and $T^{\mu \nu}$. This excess of dynamical variables comes from the additional variables introduced: $\alpha$, $\beta$, and $u^\mu$ (the thermodynamic pressure does not enter this list, since it is specified by an equation of state). As already explained, the matching conditions will provide additional constraints that will resolve this over-determination of variables.

Inserting the tensor decomposition \eqref{eq:decompos-numu-tmunu2} into the continuity equations \eqref{eq:consv-eqns}, one obtains the following dynamical equations
\begin{subequations}
\begin{align}
 \label{eq:hydro-EoM-n}
 \dot{n}_{0}+\dot{\delta n} + (n_{0}+\delta n) \theta + \partial_{\mu} \nu^{\mu} &= 0, \\
\label{eq:hydro-EoM-eps}
 \dot{\varepsilon}_{0}+\dot{\delta \varepsilon} + (\varepsilon_{0}+\delta \varepsilon + P_{0} + \Pi) \theta - \pi^{\mu \nu} \sigma_{\mu \nu} + \partial_{\mu}h^{\mu} + u_{\mu} \dot{h}^{\mu} &= 0, \\
\label{eq:hydro-EoM-umu}
(\varepsilon_{0} + \delta \varepsilon + P_{0} + \Pi)\dot{u}^{\mu} - \nabla^{\mu}(P_{0} + \Pi) + h^{\mu} \theta + h^{\alpha} \Delta^{\mu \nu} \partial_{\alpha}u_{\nu} +  \Delta^{\mu \nu} \dot{h}_{\nu} + \Delta^{\mu \nu} \partial_{\alpha}\pi^{\alpha}_{ \ \nu} &= 0.
\end{align}
\end{subequations}
Above, we defined the expansion rate $\theta \equiv \partial_{\mu}u^{\mu}$, the shear tensor $\sigma^{\mu \nu} \equiv \Delta^{\mu\nu}_{\alpha\beta}\partial^{\alpha} u^{\beta}$, the comoving time derivative $\dot{A} \equiv u^{\mu}\partial_{\mu}A$, and the space-like gradient $\nabla^{\mu} \equiv \Delta^{\mu\nu}\partial_{\nu}$. In order to close these equations, in addition to matching conditions, we must also provide constitutive or dynamical equations for all the dissipative currents: $\delta n$, $\delta \varepsilon$, $\Pi$, $\nu^\mu$, $h^\mu$, and $\pi^{\mu\nu}$. In the following sections, we discuss how to determine these novel fluid-dynamical equations of motion in a kinetic description, based on the relativistic Boltzmann equation.

\section{Boltzmann equation and fluid-dynamical variables}
\label{sec:2-BEq-fluid-vars}

The Boltzmann equation is an equation of motion for the single-particle momentum distribution function, $f(\textbf{x},\textbf{p}) \equiv f_{\textbf{p}}$. Here, we consider the relativistic Boltzmann equation for identical classical particles undergoing 2-to-2 elastic collisions,
\begin{equation}
\label{eq:EdBoltzmann}
p^{\mu }\partial_{\mu }f_{\mathbf{p}} = \int dQ \ dQ^{\prime} \ dP^{\prime} W_{pp' \leftrightarrow qq'} (f_{\mathbf{p}}f_{\mathbf{p}'}  - f_{\mathbf{q}}f_{\mathbf{q}'} )   \equiv C\left[ f_{\mathbf{p}}\right],  
\end{equation}
with $p^{\mu }=( \sqrt{m^2+\vert\textbf{p}\vert^2},\textbf{p})$ being the 4-momentum and $m$ the mass of the particles. We also introduced the Lorentz invariant transition rate, $W_{pp' \leftrightarrow qq'}$. Since we are dealing with elastic collisions, the particle number will be exceptionally conserved and will substitute the net-charge. Therefore, $N^\mu$ will now be used to refer to the particle 4-current, with $n$ being now identified as the particle number density in the local rest frame of the fluid and $\nu^{\mu}$ as the particle diffusion 4-current.

The particle 4-current and the energy-momentum tensor are determined in terms of $f_{\textbf{p}}$ as 
\cite{deGroot:80relativistic},
\begin{equation}
\label{eq:currents_kin}
\begin{aligned}
N^{\mu} &= \int dP \, p^{\mu} f_{\textbf{p}} , \\
T^{\mu\nu} &= \int dP \, p^{\mu}p^{\nu} f_{\textbf{p}} . \\
\end{aligned}
\end{equation}
Fundamental properties of the collision term guarantee that $N^{\mu}$ and $T^{\mu\nu}$ satisfy the conservation laws \eqref{eq:consv-eqns} \cite{deGroot:80relativistic,cercignani:02relativistic}. Using these identities and Eqs.\ \eqref{eq:decompos-numu-tmunu} and \eqref{eq:definitions}, we express the fluid-dynamical fields in terms of the single-particle momentum distribution function,
\begin{equation}
\begin{aligned}
\label{eq:def_kinetic}
    n &\equiv \int dP \, E_{\textbf{p}} f_{\textbf{p}} , \, \, \varepsilon \equiv \int dP \, E_{\textbf{p}}^2 f_{\textbf{p}}, \, \, P \equiv -\frac{1}{3} \int dP \, \Delta_{\mu \nu}p^{\mu}p^{\nu} f_{\textbf{p}},  \\
    \nu^{\mu} &\equiv \int dP \, p^{\langle \mu \rangle} f_{\textbf{p}}, \, \, h^{\mu} \equiv \int dP \, E_{\textbf{p}} p^{\langle \mu \rangle} f_{\textbf{p}}, \, \, \pi^{\mu\nu} \equiv \int dP \, p^{\langle \mu}p^{\nu \rangle} f_{\textbf{p}} . \\
\end{aligned}
\end{equation}
where $E_{\mathbf{p}} \equiv u_{\mu} p^{\mu}$ is the energy of the particle in the local rest frame of the system and $p^{\langle \mu \rangle} \equiv \Delta^{\mu}_{\nu} p^{\nu}$ is the projected 4-momentum. 

We now consider the fictitious equilibrium state introduced in the previous section and introduce the local equilibrium single-particle distribution function for classical particles,    
\begin{equation}
\label{eq:f_0}
    f_{0\textbf{p}} \equiv \exp{ \left(\alpha -\beta E_{\mathbf{p}} \right)} . 
\end{equation}
For the sake of convenience, we further define the non-equilibrium component of $f_{\textbf{p}}$, 
\begin{equation}
\label{eq:df}
    \delta f_{\textbf{p}} \equiv f_{\textbf{p}}-f_{0\textbf{p}} \equiv f_{0\textbf{p}} \phi_{\textbf{p}}  , 
\end{equation}
quantified in terms of the new variable $\phi_{\textbf{p}}$.
We then define the remaining fluid-dynamical variables in the following way,
\begin{equation}
\begin{aligned}
\label{eq:def_kinetic2}
    \delta n &\equiv \int dP \, E_{\textbf{p}} \delta f_{\textbf{p}}, \, \, \delta \varepsilon \equiv \int dP \, E_{\textbf{p}}^2 \delta f_{\textbf{p}}, \, \,  
    \Pi \equiv -\frac{1}{3} \int dP \, \Delta_{\mu \nu}p^{\mu}p^{\nu}\delta f_{\textbf{p}}. 
\end{aligned}
\end{equation}

We now discuss the possible matching conditions that can be employed to define the local equilibrium fields, $T$, $\mu$, and $u^{\mu}$. First, we note that the Landau matching conditions, given in Eqs.\ \eqref{eq:Landau} and \eqref{eq:decompos-numu-tmunu2}, can be expressed as the following 5 constrains for $\delta f_{\textbf{p}}$, 
\begin{equation}
\begin{aligned}
\label{eq:Landau_kinetic}
    \langle E_{\textbf{p}} \phi_{\textbf{p}} \rangle_{0} = 0, \, \, 
    \langle E_{\textbf{p}}^{2} \phi_{\textbf{p}} \rangle_{0} = 0, \, \,
    \left\langle E_{\textbf{p}} p^{\langle \mu \rangle} \phi_{\textbf{p}} \right\rangle_{0} =0,\\
\end{aligned}
\end{equation}
where here we introduced the notation for integrals over the local equilibrium distribution, $\langle \cdots \rangle_0 = \int dP (\cdots) f_{0\textbf{p}}$. On the other hand, the Eckart matching conditions provide the following constraints, 
\begin{equation}
\begin{aligned}
\label{eq:Eckart_kinetic}
  \langle E_{\textbf{p}} \phi_{\textbf{p}} \rangle_{0} = 0, \, \, 
    \langle E_{\textbf{p}}^{2} \phi_{\textbf{p}} \rangle_{0} = 0, \, \,
    \left\langle  p^{\langle \mu \rangle} \phi_{\textbf{p}} \right\rangle_{0} =0,\\
\end{aligned}
\end{equation}
with only the last constraint being modified, in comparison with Landau's matching conditions. In kinetic theory, these set of constraints can be generalized in the following way
\begin{equation}
\begin{aligned}
\label{eq:matching_kinetic1}
    \left\langle \, g_{\textbf{p}}  \phi_{\textbf{p}} \right\rangle = 0, \ 
    \left\langle \, h_{\textbf{p}}  \phi_{\textbf{p}} \right\rangle = 0, \
    \left\langle \, q_{\textbf{p}} p^{\langle \mu \rangle} \phi_{\textbf{p}} \right\rangle =0, \\
\end{aligned}
\end{equation}
where $g_{\textbf{p}}$ and $h_{\textbf{p}}$ are linearly independent functions of $E_{\textbf{p}}$ and $q_{\textbf{p}}$ is an arbitrary function of $E_{\textbf{p}}$. Once these functions are specified, one determines the matching condition being employed. In general, we see that these matching conditions are imposed using moments of $\delta f_{\textbf{p}}$ that are not contained in any conserved current. The Eckart and Landau pictures correspond to exceptional cases, in which we actually define the local equilibrium state using components of $N^{\mu}$ and $T^{\mu \nu}$. In this work, we shall simplify the general conditions \eqref{eq:matching_kinetic1} by assuming the following form for the functions $g_{\textbf{p}}$, $h_{\textbf{p}}$, and $q_{\textbf{p}}$,
\begin{equation}
\begin{aligned}
\label{eq:matching_kinetic2}
    g_{\textbf{p}} = E_{\textbf{p}}^q, \ 
    h_{\textbf{p}} = E_{\textbf{p}}^s, \ 
    q_{\textbf{p}} = E_{\textbf{p}}^z, \ 
\end{aligned}
\end{equation}
with $q \neq s$, which include Landau and Eckart matching conditions as the particular cases where $(q,s,z) =(1,2,0)$ and $(q,s,z) = (1,2,1)$, respectively. In the next section, we will derive equations of motion for the relevant irreducible moments of the single particle distribution function $f_{\mathbf{p}}$.

\section{Moment equations}
\label{sec:3-Moment-eqs}

In this work, the equations of motion for the dissipative components of $N^{\mu}$ and $T^{\mu \nu}$ will be obtained using the method of moments. In this approach, one expands the single-particle momentum distribution function in momentum space using a complete and orthogonal basis \cite{denicol2012derivation}. Such procedure expresses $f(\mathbf{x},\mathbf{p})$ in terms of its irreducible moments
\begin{equation}
\label{eq:irreducible_moments}
\begin{aligned}
\rho^{\mu_{1} \cdots \mu_{\ell}}_{r} = \left\langle  E_{\mathbf{p}}^{r} p^{\langle \mu_{1}} \cdots p^{\mu_{\ell} \rangle} \phi_{\mathbf{p}} \right\rangle_{0} , 
\end{aligned}    
\end{equation}
where $\phi_{\mathbf{p}}$ is given in Eq.\ \eqref{eq:df} and we used the irreducible tensors $p^{\langle \mu_{1}} \cdots p^{\mu_{\ell} \rangle}$, defined from the following projection of the tensor $p^{\mu_{1}} \cdots p^{\mu_{\ell}}$,  $A^{\langle \mu_{1} \cdots \mu_{\ell}\rangle} \equiv \Delta^{\mu_{1} \cdots \mu_{\ell}}_{\nu_{1} \cdots \nu_{\ell}} A^{\nu_{1} \cdots \nu_{\ell}}$. Here, $\Delta^{\mu_{1} \cdots \mu_{\ell}}_{\nu_{1} \cdots \nu_{\ell}}$ is a $2\ell$--rank tensor, constructed solely using the projection operators $\Delta^{\mu}_{\nu}$, in such a way that it is symmetric under the exchange of the indices ($\mu_{1} \cdots \mu_{\ell}$) and ($\nu_{1} \cdots \nu_{\ell}$), separately, and also traceless in each subset of indices. These projection operators were formally defined and constructed in Ref.\ \cite{deGroot:80relativistic} and used extensively in Ref.\ \cite{denicol2012derivation}.   

Therefore, in the method of moments the problem of solving for the single-particle distribution function  $f(\mathbf{x},\mathbf{p})$ using the Boltzmann equation is converted into solving an infinite tower of coupled differential equations for its irreducible moments. These equations of motion were originally derived in Ref.\ \cite{denicol2012derivation}, for the irreducible moments of rank 0, 1, and 2, assuming Landau matching conditions. In this paper, we re-derive these moment equations following the procedure outlined in \cite{denicol2012derivation}, but for an arbitrary matching condition. We then obtain the following equations for the scalar moments,
\begin{equation}
\label{eq:transient-l=0}
\begin{aligned}
& \dot{\rho}_{r} 
- 
r \dot{u}_{\mu} \rho_{r-1}^{\mu} 
+
\nabla_{\mu} \rho^{\mu}_{r-1}
+  \frac{\theta}{3} \left[ -m^{2} (r-1) \rho_{r-2} +(r+2) \rho_{r} \right] - (r-1) \rho_{r-2}^{\mu \nu} \sigma_{\mu \nu}\\
& 
+
\left(- \frac{I_{30}}{D_{20}}I_{r0}  +  \frac{I_{20}}{D_{20}} I_{r+1,0} \right) \left( \dot{\delta n} 
+
\delta n \ \theta
+
\partial_{\mu} \nu^{\mu} \right) 
 \\ & 
+
\left( \frac{I_{20}}{D_{20}} I_{r0} - \frac{I_{10}}{D_{20}} I_{r+1,0} \right) \left( \delta \varepsilon \ \theta
+
\Pi \theta  
-
\pi^{\alpha \beta} \sigma_{\alpha \beta} 
+
\partial_{\mu} h^{\mu}
+
u_{\mu} \dot{h}^{\mu}
+ 
 \dot{\delta \varepsilon} \right)
\\
&
+
\left[ - \frac{n_{0}}{D_{20}}  G_{3r} + \frac{\varepsilon_{0} + P_{0}}{D_{20}}G_{2r} + (r-1) I_{r1} + I_{r0} \right]  \theta 
=
\left\langle E_{\mathbf{p}}^{r-1} C[f_{\mathbf{p}}] \right\rangle_{0},
\end{aligned}    
\end{equation}
the following equation for the irreducible rank-1 moments,
\begin{equation}
\label{eq:transient-l=1}
\begin{aligned}
& 
\dot{\rho}_{r}^{\langle \alpha \rangle} 
-
r \dot{u}_{\mu} \rho^{\mu \alpha}_{r-1} 
+
\frac{1}{3} \dot{u}^{\alpha} \left[ -  m^{2} r \rho_{r-1} + (r+3) \rho_{r+1} \right]
+
\omega_{\mu}^{\ \alpha} \rho^{\mu}_{r}
+
\Delta^{\alpha}_{ \ \alpha'} \nabla_{\mu} \rho_{r-1}^{\alpha' \mu} 
\\
&
-
(r-1) \sigma_{\mu \nu} \rho^{\mu \alpha \nu}_{r-2}
+
\frac{1}{3} \nabla^{\alpha}\left( m^{2} \rho_{r-1} - \rho_{r+1} \right)
+
\frac{\theta}{3} \left[ (r+3) \rho_{r}^{\alpha} - (r-1) m^{2} \rho_{r-2}^{\alpha}\right]
\\
& 
+
\frac{1}{5} \sigma_{\mu}^{\ \alpha} \left[ (2r+3) \rho_{r}^{\mu} - 2(r-1) m^{2}\rho_{r-2}^{\mu}\right] \\
&
- 
\beta
 \frac{ I_{r+2,1} }{\varepsilon_{0} +P_{0}} \left[ \Delta^{\alpha}_{ \ \alpha'} \partial_{\nu} \pi^{\nu \alpha'}
+
\Delta^{\alpha \mu} \dot{h}_{\mu}
+
h^{\beta} \sigma_{\beta}^{\ \alpha}
+
h^{\beta} \omega_{\beta}^{\ \alpha}
+
\frac{4}{3}
 h^{\alpha} \theta
-
\nabla^{\alpha} \Pi 
+
(\delta \varepsilon+ \Pi) \dot{u}^{\alpha} \right]
\\ 
& 
- 
 \left(I_{r+1,1} - \frac{n I_{r+2,1}}{\varepsilon_{0} + P_{0}}\right) \nabla^{\alpha} \alpha 
=
\left\langle E_{\mathbf{p}}^{r-1} p^{\langle \alpha \rangle} C[f_{\mathbf{p}}] \right\rangle_{0},
\end{aligned}    
\end{equation}
and the following equations for the irreducible rank-2 moments,
\begin{equation}
\label{eq:transient-l=2}
\begin{aligned}
&  \dot{\rho}^{\langle \alpha \beta \rangle}_{r}
-
r \dot{u}_{\mu} \rho^{\mu \alpha \beta}_{r-1} 
+
\frac{2}{5}  \dot{u}^{ \langle \alpha} \left[ \left( r + 5\right) \rho_{r+1}^{\beta\rangle} - r m^{2} \rho_{r-1}^{\beta\rangle} \right]
+\Delta^{\alpha \beta}_{ \ \ \alpha' \beta'}
 \nabla_{\mu} \rho_{r-1}^{\alpha' \beta' \mu} 
\\
& 
-
(r-1) \sigma_{\mu \nu} \rho^{\mu \alpha \beta \nu}_{r-2}
+
\frac{2}{5} 
 \nabla^{\langle \alpha}\left( m^{2} \rho_{r-1}^{\beta \rangle} - \rho_{r+1}^{\beta \rangle} \right)
+
\frac{\theta}{3} \left[ (r+4) \rho_{r}^{\alpha \beta} - (r-1) m^{2} \rho_{r-2}^{\alpha \beta}\right]
+
2 
 \omega_{\mu}^{ \ \langle \alpha \vert} \rho^{\mu \vert \beta \rangle}_{r} 
\\
& 
+ \frac{2}{7} 
\sigma_{\mu}^{\ \langle \alpha}  \left[ (2r+5) \rho_{r}^{\beta \rangle \mu} - 2(r-1) m^{2}\rho_{r-2}^{\beta \rangle \mu}\right] 
+
\frac{2}{15} \sigma^{\alpha \beta}  \left[ - (r-1)m^{4} \rho_{r-2} + (2r+3) m^{2}\rho_{r} - (r+4) \rho_{r+2}\right]\\
&
-2 I_{r+3,2}
 \beta \sigma^{\alpha \beta}
=
\left\langle E_{\mathbf{p}}^{r-1} p^{\langle \alpha} p^{\beta \rangle} C[f_{\mathbf{p}}] \right\rangle .
\end{aligned}    
\end{equation}
Above, we defined the vorticity tensor $\omega^{\mu \nu} \equiv  \left( \partial^{\mu} u^{\nu} - \partial^{\nu} u^{\mu} \right)/2$ and also made use of the thermodynamic functions \cite{denicol2012derivation}
\begin{equation}
\begin{aligned}
& I_{nq} = \frac{1}{(2q+1)!!} \left\langle \left( -\Delta^{\lambda \sigma}p_{\lambda} p_{\sigma} \right)^{q} E_{\mathbf{p}}^{n-2q}\right\rangle_{0},
\, \,
D_{nq} = I_{n+1,q} I_{n-1,q} - I_{nq}^{2} ,\\
&
G_{nm} = I_{n0} I_{m0} - I_{n-1,0}I_{m+1,0}
.
\end{aligned}    
\end{equation}
We further used the identity
\begin{equation}
\begin{aligned}
\label{eq:Gibbs-duhem}
\nabla_{\mu} \beta = \frac{1}{\varepsilon_{0} + P_{0}} \left( n_{0} \nabla_{\mu} \alpha  - \beta \nabla_{\mu} P_{0} \right),
\end{aligned}
\end{equation}
which stems from the Gibbs-Duhem relation, to obtain the last term of the left-hand side of Eq.\ \eqref{eq:transient-l=1} and used the notation
\begin{equation}
\begin{aligned}
& \dot{\rho}_{r}^{\langle \mu_{1} \cdots \mu_{\ell} \rangle} = \Delta^{\mu_{1} \cdots \mu_{\ell}}_{\nu_{1} \cdots \nu_{\ell}} \dot{\rho}_{r}^{\nu_{1} \cdots \nu_{\ell}} .
\end{aligned}    
\end{equation}
The above equations form a subset of an infinite tower of coupled differential equations mixing different irreducible moments of various ranks. The subset of moment equations shown above (of rank 0, 1, and 2) are the ones that are actually relevant to the derivation of fluid-dynamical equations, as originally shown in Ref.\ \cite{denicol2012derivation}.

We note that some of the irreducible moments defined in Eq.\ \eqref{eq:irreducible_moments} can be identified with the fluid-dynamical fields introduced in the previous section. Namely, those are
\begin{equation}
\label{eq:moments-and-hydro}
\begin{aligned}
& \delta n = \rho_{1}, \, \, \,
\delta \varepsilon = \rho_{2} , \, \, \,
\Pi = \frac{1}{3} (\rho_{2} - m^{2} \rho_{0}) ,\\
& \nu^{\alpha} = \rho_{0}^{\alpha}, \, \, \, 
h^{\alpha} = \rho_{1}^{\alpha}, \, \, \,
\pi^{\alpha \beta} = \rho_{0}^{\alpha \beta}. \\
\end{aligned}    
\end{equation}
Moreover, the general matching and frame conditions given by Eqs.\ \eqref{eq:matching_kinetic1} and \eqref{eq:matching_kinetic2} can be written in a simple way using this notation,  
\begin{subequations}
\begin{align}
\rho_{q} & = \rho_{s} = 0, \, \,  (q \neq s) \label{eq:matching-conditions_a} \\ \rho_{z}^{\mu} & = 0. \label{eq:matching-conditions_b}  
\end{align}    
\end{subequations}
For $(q,s,z) = (1,2,0)$, one has Landau matching conditions. Meanwhile, taking $(q,s,z) = (1,2,1)$ one imposes Eckart's matching conditions. As we can see, more general matching conditions lead to constraints that involve non-fluid-dynamic fields. 

Finally, we note that the moment equations derived above are not closed nor expressed in terms of fluid-dynamical variables. In the next section we proceed to derive fluid-dynamical equations of motion from the moment equations above by generalizing the well-known 14-moment approximation \cite{israel1979annals} to be consistent with arbitrary matching conditions. We shall refer to this new procedure as 19-moment approximation, since fluid-dynamical approximations for arbitrary matching conditions involve a total of 19 fields, instead of the usual 14. 

\section{Truncation scheme: 19 moments approximation}
\label{sec:4-truncation-19}

Our objective in this section is to derive a system of equations involving only the 19 fluid-dynamic fields $(n, \delta n, \varepsilon, \delta \varepsilon, \Pi , u^{\mu}, \nu^{\mu}, h^{\mu}, \pi^{\mu \nu})$ from Eqs.\  \eqref{eq:transient-l=0}-\eqref{eq:transient-l=2}. Here, we shall follow the idea originally proposed by H.\ Grad \cite{grad:1949kinetic}, in the non-relativistic regime, and later updated by Israel and Stewart \cite{israel1979annals} to the relativistic regime, and close the moment equations by an explicit truncation of the momentum expansion of $f_{\mathbf{p}}$. 

We expand the non-equilibrium correction $\phi_{\mathbf{p}}$ in momentum using a complete basis made of irreducible tensors, $p^{\langle \mu_{1}} \cdots p^{\mu_{\ell} \rangle}$ , and orthogonal polynomials, $P^{(\ell)}_{n}$, constructed using powers of $E_{\mathbf{p}}$ in such a way that $P^{(\ell)}_{0} \equiv 1$. This leads to \cite{denicol2012derivation},
\begin{equation}
\label{eq:moment_exp}
\phi_{\mathbf{p}} = \sum_{\ell,n = 0}^{\infty}  \Phi_{n}^{\mu_{1} \cdots \mu_{\ell}} P_{n}^{(\ell)} p_{\langle \mu_{1}} \cdots  p_{\mu_{\ell} \rangle}.
\end{equation}
The basis elements satisfy the following orthogonality conditions, 
\begin{equation}
\label{eq:polys-orthogonal}
\begin{aligned}
\int dP
p^{\langle \mu_{1}} \cdots p^{\mu_{\ell} \rangle}
p_{\langle \nu_{1}} \cdots p_{\nu_{m} \rangle}
H(E_{\mathbf{p}})
& = \frac{\ell!\delta_{\ell m}}{(2\ell + 1)!!}  \Delta^{\mu_{1} \cdots \mu_{\ell}}_{\nu_{1} \cdots \nu_{\ell}} \int dP \left(\Delta^{\mu \nu} p_{\mu} p_{\nu} \right)^{\ell} H(E_{\mathbf{p}}),  \\
\left( P^{(\ell)}_{m} ,  P^{(\ell)}_{n}   \right)_{\ell} 
&= A^{(\ell)}_{m} \delta_{mn}, 
\end{aligned}    
\end{equation}
 where $H(E_{\mathbf{p}})$ is an arbitrary function of $E_{\mathbf{p}}$ and, for the second orthogonality relation, we used the notation
\begin{equation}
\label{eq:inner-prod-l}
\begin{aligned}
\left( \phi , \psi \right)_{\ell} 
= 
\frac{\ell!}{(2\ell + 1)!!} \left\langle \left( \Delta^{\mu \nu} p_{\mu} p_{\nu} \right)^{\ell}   \phi_{\mathbf{p}} \psi_{\mathbf{p}} \right\rangle_{0}.
\end{aligned}    
\end{equation}

These orthogonality conditions allow us to express the expansion coefficients $\Phi_{n}^{\mu_{1} \cdots \mu_{\ell}}$ in terms of the irreducible moments defined in Eq.\ \eqref{eq:irreducible_moments}. However, this will not be required here since we shall follow the procedure originally proposed by Israel and Stewart and truncate the moment expansion \eqref{eq:moment_exp} by hand, in such a way that only a finite number of expansion coefficients remain (in Israel-Stewart's original approach, the truncation is performed in such a way that 14 expansion coefficients remain). The remaining expansion coefficients will then be matched to the fluid-dynamical fields. The truncated expansion is,
\begin{equation}
\label{eq:truncation-Phi's}
\begin{aligned}
& 
\phi_{\mathbf{p}} \simeq \sum_{n=0}^{N_{0}} \Phi_{n} P^{(0)}_{n} 
+
\sum_{n=0}^{N_{1}} \Phi_{n}^{\mu} P^{(1)}_{n} p_{\langle \mu \rangle} 
+
\Phi_{0}^{\mu \nu}  p_{\langle \mu} p_{\nu \rangle}.
\end{aligned}
\end{equation}
We shall always set $N_{2}=1$. On the other hand the truncation parameters $N_{0}$ and $N_{1}$ will depend on the matching condition employed. This happens because the matching conditions determine the number of independent scalar and rank-1 fields that appear in the fluid-dynamical formulation. The maximum number of degrees of freedom that can occur in fluid-dynamical theories with arbitrary matching conditions is 19 and, in such cases, we set $(N_{0},N_{1}) = (4,2)$ in order to enforce that we have 5 scalar fields, 3 vector fields, and one second-rank traceless tensor. Note that if we were employing Landau matching conditions, the same truncation is performed with $(N_{0},N_{1}) = (2,1)$, leading to 14 independent degrees of freedom (since $\delta n$, $\delta \varepsilon$ and $h^\mu$ are set to zero in this matching procedure). By construction, in this truncation scheme all irreducible moments, $\rho_{r}^{\mu_1 \cdots \mu_\ell}$, of rank higher than two vanish. The irreducible moments of rank 0, 1, and 2 will always be expressed in terms of the expansion coefficients $\Phi_n$, $\Phi_{n}^{\mu}$, and $\Phi_{n}^{\mu \nu}$.     

The next step is to determine the expansion coefficients using the definitions of the dissipative currents, given in Eqs.\ \eqref{eq:def_kinetic} and \eqref{eq:def_kinetic2}, and the general matching conditions provided in Eqs.\ \eqref{eq:matching-conditions_a} and \eqref{eq:matching-conditions_b}. This leads to a system of algebraic equations that can be solved separately for the scalar, vector and tensor expansion coefficients. The expression for the 2-rank expansion coefficient, $\Phi_{0}^{\mu \nu}$, is obtained by inserting the truncated moment expansion \eqref{eq:truncation-Phi's} in the definition of the shear stress tensor given in Eq.\ \eqref{eq:def_kinetic}. This leads to the simple relation,
\begin{equation}
\label{eq:eqs-Phi's-l=2}
\Phi^{\mu \nu}_{0} = \frac{\pi^{\mu \nu}}{ (1,1)_2} = \frac{\pi^{\mu \nu}}{2 I_{42}} .
\end{equation}
The equations for the vector and scalar expansion coefficients are more complicated and will be discussed in the next two subsections.

\subsection{Vector components}
\label{subs:vec-comp-Phis}

Equations for the vector expansion coefficients, $\Phi_{0}^{\mu}$, $\Phi_{1}^{\mu}$, and $\Phi_{2}^{\mu}$, are obtained by inserting the truncated moment expansion \eqref{eq:truncation-Phi's} into the definitions of $\nu^{\mu}$ and $h^{\mu}$, given in Eq.\ \eqref{eq:def_kinetic}, and also into the matching condition \eqref{eq:matching-conditions_b}. This will lead to three distinct equations that can be cast in the following matrix form,
\begin{equation}
\label{eq:eqs-Phi's-l=1}
\begin{aligned}
&\left( \begin{array}{ccc}
 (1, 1)_{1} & 0 & 0 \\
 (P^{(1)}_{0}, E_{\mathbf{p}})_{1} & (P^{(1)}_{1}, E_{\mathbf{p}})_{1}  & 0 \\
(P^{(1)}_{0}, E_{\mathbf{p}}^{z})_{1} & (P^{(1)}_{1}, E_{\mathbf{p}}^{z})_{1}  & (P^{(1)}_{2}, E_{\mathbf{p}}^{z})_{1} 
\end{array}
\right)
\left(
\begin{array}{cc}
  \Phi^{\mu}_{0}     \\
  \Phi^{\mu}_{1}     \\
  \Phi^{\mu}_{2}
\end{array}
\right)
=
\left(\begin{array}{cc}
\nu^{\mu}       \\
h^{\mu}       \\
0       
\end{array}
\right).
\end{aligned}    
\end{equation}
Thus, once this equation is inverted, $\Phi_{0}^{\mu}$, $\Phi_{1}^{\mu}$, and $\Phi_{2}^{\mu}$, will be expressed as a linear combination of $\nu^{\mu}$ and $h^{\mu}$. The final inverted expressions are given in Appendix \ref{apn:approx-moms}.  

It is important to be careful when applying this inversion procedure using either the Eckart ($z=0$) or the Landau ($z=1$) matching conditions. In the case of Eckart's picture, the first and third equations depicted (in matrix form) above become identical and, thus, one of them has to be removed. Similarly, in the case of Landau's picture, the second and third equations depicted above become identical and one of those has to be removed. This effectively corresponds to modifying the truncation of the moment expansion \eqref{eq:truncation-Phi's} by taking $N_1 = 1$. This new procedure, specific to these two matching conditions, will lead to the following equation for $\Phi_{0}^{\mu}$ and $\Phi_{1}^{\mu}$:
\begin{itemize}
    \item Eckart Matching Conditions
\begin{equation}
\label{eq:eqs-Phi's-l=1-Eckart}
\begin{aligned}
&\left( \begin{array}{cc}
(1,1)_1 & 0  \\
(P^{(1)}_{0}, E_{\mathbf{p}})_{1} & (P^{(1)}_{1}, E_{\mathbf{p}})_{1}
\end{array}
\right)
\left(
\begin{array}{cc}
  \Phi^{\mu}_{0}     \\
  \Phi^{\mu}_{1}     \\
\end{array}
\right)
=
\left(\begin{array}{cc}
0       \\
h^{\mu}       
\end{array}
\right).
\end{aligned}    
\end{equation}
  \item Landau Matching Conditions
\begin{equation}
\label{eq:eqs-Phi's-l=1-Landau}
\begin{aligned}
&\left( \begin{array}{cc}
 (1, 1)_{1} & 0 \\
 (P^{(1)}_{0}, E_{\mathbf{p}})_{1} & (P^{(1)}_{1}, E_{\mathbf{p}})_{1}  
\end{array}
\right)
\left(
\begin{array}{cc}
  \Phi^{\mu}_{0}     \\
  \Phi^{\mu}_{1}     \\
\end{array}
\right)
=
\left(\begin{array}{cc}
\nu^{\mu}       \\
0       
\end{array}
\right).
\end{aligned}    
\end{equation}
\end{itemize}

\subsection{Scalar components}
\label{subs:sca-comp-Phis}

Equations for the scalar expansion coefficients, $\Phi_{0}$, $\Phi_{1}$, $\Phi_{2}$, $\Phi_{3}$, and $\Phi_{4}$ are obtained by inserting the truncated moment expansion \eqref{eq:truncation-Phi's} into the definitions of $\delta n$, $\delta \varepsilon$  and $\Pi$, given in Eq.\ \eqref{eq:def_kinetic}, and also into the matching condition \eqref{eq:matching-conditions_a}. This will lead to five distinct equations that can be cast in the following matrix form,
\begin{equation}
\label{eq:eqs-Phi's-l=0}
\begin{aligned}
\left( \begin{array}{ccccc}
 \left( P^{(0)}_{0} , E_{\mathbf{p}} \right)_{0} & \left( P^{(0)}_{1}, E_{\mathbf{p}} \right)_{0} & \left( P^{(0)}_{2}, E_{\mathbf{p}} \right)_{0} & \left( P^{(0)}_{3}, E_{\mathbf{p}} \right)_{0} & \left( P^{(0)}_{4}, E_{\mathbf{p}} \right)_{0} \\
 \left( P^{(0)}_{0} , E_{\mathbf{p}}^{2} \right)_{0} & \left( P^{(0)}_{1}, E_{\mathbf{p}}^{2} \right)_{0} & \left( P^{(0)}_{2} , E_{\mathbf{p}}^{2} \right)_{0} & \left( P^{(0)}_{3}, E_{\mathbf{p}}^{2} \right)_{0} & \left( P^{(0)}_{4}, E_{\mathbf{p}}^{2} \right)_{0} \\
 \left( P^{(0)}_{0} , E_{\mathbf{p}}^{2} - m^{2} \right)_{0}    & \left( P^{(0)}_{1}, E_{\mathbf{p}}^{2} - m^{2} \right)_{0} & \left( P^{(0)}_{2} , E_{\mathbf{p}}^{2} - m^{2} \right)_{0} & \left( P^{(0)}_{3}, E_{\mathbf{p}}^{2} - m^{2}\right)_{0} & \left( P^{(0)}_{4}, E_{\mathbf{p}}^{2} - m^{2} \right)_{0} \\
 \left( P^{(0)}_{0} , E_{\mathbf{p}}^{q} \right)_{0} & \left( P^{(0)}_{1}, E_{\mathbf{p}}^{q} \right)_{0} & \left( P^{(0)}_{2} , E_{\mathbf{p}}^{q} \right)_{0} & \left( P^{(0)}_{3} , E_{\mathbf{p}}^{q} \right)_{0} & \left( P^{(0)}_{4}, E_{\mathbf{p}}^{q} \right)_{0} \\
 \left( P^{(0)}_{0} , E_{\mathbf{p}}^{s} \right)_{0} & \left( P^{(0)}_{1}, E_{\mathbf{p}}^{s} \right)_{0} & \left( P^{(0)}_{2} , E_{\mathbf{p}}^{s} \right)_{0} & \left( P^{(0)}_{3} , E_{\mathbf{p}}^{s} \right)_{0} & \left( P^{(0)}_{4}, E_{\mathbf{p}}^{s} \right)_{0}
 \end{array}\right)
\left( \begin{array}{c}
\Phi_{0} \\
\Phi_{1} \\
\Phi_{2} \\
\Phi_{3} \\
\Phi_{4} \\
\end{array}\right)
=
\left( \begin{array}{c}
\delta n \\
\delta \varepsilon \\
3 \Pi \\
0 \\
0  
\end{array}\right).
\end{aligned}    
\end{equation}
Thus, once this equation is inverted,
$\Phi_{0}$, $\Phi_{1}$, $\Phi_{2}$, $\Phi_{3}$, and $\Phi_{4}$ will be expressed as a linear combination of $\delta n$, $\delta \varepsilon$,  and $\Pi$. The final inverted expressions are given in Appendix \ref{apn:approx-moms}.  

Similarly to what occurred when obtaining the vector expansion coefficients, it is important to be careful when applying this inversion procedure using either the Eckart or the Landau ($q=1$, $s=2$) matching conditions. In this case, the first and second equations depicted (in matrix form) above become identical to the last two equations, thus, the last two should be removed. This is implemented by modifying the truncation of the moment expansion \eqref{eq:truncation-Phi's} by taking $N_0 = 2$. This new procedure, specific to these two matching conditions, will lead to the following equation for $\Phi_{0}$, $\Phi_{1}$, and $\Phi_{2}$:
\begin{itemize}
    \item Landau or Eckart matching conditions  ($q=1$, $s=2$)
\begin{equation}
\label{eq:eqs-Phi's-l=0-}
\begin{aligned}
\left( \begin{array}{ccccc}
 \left( P^{(0)}_{0} , E_{\mathbf{p}} \right)_{0} & \left( P^{(0)}_{1}, E_{\mathbf{p}} \right)_{0} & \left( P^{(0)}_{2}, E_{\mathbf{p}} \right)_{0} \\
 \left( P^{(0)}_{0} , E_{\mathbf{p}}^{2} \right)_{0} & \left( P^{(0)}_{1}, E_{\mathbf{p}}^{2} \right)_{0} & \left( P^{(0)}_{2}, E_{\mathbf{p}}^{2} \right)_{0} \\
 \left( P^{(0)}_{0} , E_{\mathbf{p}}^{2} - m^{2} \right)_{0}    & \left( P^{(0)}_{1}, E_{\mathbf{p}}^{2} - m^{2} \right)_{0} & \left( P^{(0)}_{2}, E_{\mathbf{p}}^{2} - m^{2} \right)_{0} 
 \end{array}\right)
\left( \begin{array}{c}
\Phi_{0} \\
\Phi_{1} \\
\Phi_{2} \\
\end{array}\right)
=
\left( \begin{array}{c}
0 \\
0 \\
3 \Pi \\
\end{array}\right).
\end{aligned}    
\end{equation}
\end{itemize}

Similar problems will occur to matching conditions with either ($q$ or $s=1$) or ($q$ or $s=2$). In the first case, one of the matching conditions will lead to an equation identical to the first equation depicted in \eqref{eq:eqs-Phi's-l=0}. While, in the latter case, one of the matching conditions will lead to an equation identical to the second equation depicted in \eqref{eq:eqs-Phi's-l=0}. Thus, in each case, one of the conditions must be removed. As before, this is implemented by modifying the truncation of the moment expansion \eqref{eq:truncation-Phi's} by taking $N_0 = 3$. This new procedure, will lead to the following equation for $\Phi_{0}$, $\Phi_{1}$, $\Phi_{2}$, and $\Phi_{3}$:
\begin{itemize}
\item  Matchings with $\delta n = 0$ but $\delta \varepsilon \neq 0$ ($q=1$, $s \neq 2$)

\begin{equation}
\label{eq:eqs-Phi's-l=0---}
\begin{aligned}
\left( \begin{array}{ccccc}
 \left( P^{(0)}_{0} , E_{\mathbf{p}}^{2} \right)_{0} & \left( P^{(0)}_{1}, E_{\mathbf{p}}^{2} \right)_{0} & \left( P^{(0)}_{2} , E_{\mathbf{p}}^{2} \right)_{0} & \left( P^{(0)}_{3}, E_{\mathbf{p}}^{2} \right)_{0} \\
 \left( P^{(0)}_{0} , E_{\mathbf{p}}^{2} - m^{2} \right)_{0}    & \left( P^{(0)}_{1}, E_{\mathbf{p}}^{2} - m^{2} \right)_{0} & \left( P^{(0)}_{2} , E_{\mathbf{p}}^{2} - m^{2} \right)_{0} & \left( P^{(0)}_{3}, E_{\mathbf{p}}^{2} - m^{2}\right)_{0}  \\
 \left( P^{(0)}_{0} , E_{\mathbf{p}} \right)_{0} & \left( P^{(0)}_{1}, E_{\mathbf{p}} \right)_{0} & \left( P^{(0)}_{2} , E_{\mathbf{p}} \right)_{0} & \left( P^{(0)}_{3} , E_{\mathbf{p}} \right)_{0} \\
 \left( P^{(0)}_{0} , E_{\mathbf{p}}^{s} \right)_{0} & \left( P^{(0)}_{1}, E_{\mathbf{p}}^{s} \right)_{0} & \left( P^{(0)}_{2} , E_{\mathbf{p}}^{s} \right)_{0} & \left( P^{(0)}_{3} , E_{\mathbf{p}}^{s} \right)_{0} 
 \end{array}\right)
\left( \begin{array}{c}
\Phi_{0} \\
\Phi_{1} \\
\Phi_{2} \\
\Phi_{3} \\
\end{array}\right)
=
\left( \begin{array}{c}
\delta \varepsilon \\
3 \Pi \\
0 \\
0  
\end{array}\right).
\end{aligned}    
\end{equation}

\item  Matchings with $\delta n \neq 0$ but $\delta \varepsilon = 0$ ($q =2$, $s \neq 1$)

\begin{equation}
\label{eq:eqs-Phi's-l=0--}
\begin{aligned}
\left( \begin{array}{ccccc}
 \left( P^{(0)}_{0} , E_{\mathbf{p}} \right)_{0} & \left( P^{(0)}_{1}, E_{\mathbf{p}} \right)_{0} & \left( P^{(0)}_{2}, E_{\mathbf{p}} \right)_{0} & \left( P^{(0)}_{3}, E_{\mathbf{p}} \right)_{0}  \\
 \left( P^{(0)}_{0} , E_{\mathbf{p}}^{2} - m^{2} \right)_{0}    & \left( P^{(0)}_{1}, E_{\mathbf{p}}^{2} - m^{2} \right)_{0} & \left( P^{(0)}_{2} , E_{\mathbf{p}}^{2} - m^{2} \right)_{0} & \left( P^{(0)}_{3}, E_{\mathbf{p}}^{2} - m^{2}\right)_{0} \\
 \left( P^{(0)}_{0} , E_{\mathbf{p}}^{2} \right)_{0} & \left( P^{(0)}_{1}, E_{\mathbf{p}}^{2} \right)_{0} & \left( P^{(0)}_{2} , E_{\mathbf{p}}^{2} \right)_{0} & \left( P^{(0)}_{3} , E_{\mathbf{p}}^{2} \right)_{0} \\
 \left( P^{(0)}_{0} , E_{\mathbf{p}}^{s} \right)_{0} & \left( P^{(0)}_{1}, E_{\mathbf{p}}^{s} \right)_{0} & \left( P^{(0)}_{2} , E_{\mathbf{p}}^{s} \right)_{0} & \left( P^{(0)}_{3} , E_{\mathbf{p}}^{s} \right)_{0} 
 \end{array}\right)
\left( \begin{array}{c}
\Phi_{0} \\
\Phi_{1} \\
\Phi_{2} \\
\Phi_{3} \\
\end{array}\right)
=
\left( \begin{array}{c}
\delta n \\
3 \Pi \\
0 \\
0  
\end{array}\right).
\end{aligned}    
\end{equation}
\end{itemize}

There is yet another case to be considered, corresponding to the limit of massless particles, $m \rightarrow 0$. In this case, the energy-momentum tensor becomes traceless, $T^{\mu}_{\ \mu} = \varepsilon_{0} + \delta \varepsilon - 3P_{0} - 3\Pi = 0$, effectively removing one degree of freedom from our system. This will affect all the matching conditions displayed so far in this subsection. The matching conditions associated to bulk viscous pressure become redundant and must be removed. Also, we remove one additional scalar term from the corresponding moment expansion of $\phi_{\mathbf{p}}$. In this case, we re-write each matching procedure described above as:

\begin{itemize}
    \item  Massless limit: General frame 
\begin{equation}
\label{eq:eqs-Phi's-l=0-m-0}
\begin{aligned}
\left( \begin{array}{cccc}
 \left( P^{(0)}_{0} , E_{\mathbf{p}} \right)_{0} & \left( P^{(0)}_{1}, E_{\mathbf{p}} \right)_{0} & \left( P^{(0)}_{2}, E_{\mathbf{p}} \right)_{0} & \left( P^{(0)}_{3}, E_{\mathbf{p}} \right)_{0} \\
 \left( P^{(0)}_{0} , E_{\mathbf{p}}^{2} \right)_{0} & \left( P^{(0)}_{1}, E_{\mathbf{p}}^{2} \right)_{0} & \left( P^{(0)}_{2} , E_{\mathbf{p}}^{2} \right)_{0} & \left( P^{(0)}_{3}, E_{\mathbf{p}}^{2} \right)_{0}\\
 \left( P^{(0)}_{0} , E_{\mathbf{p}}^{q} \right)_{0} & \left( P^{(0)}_{1}, E_{\mathbf{p}}^{q} \right)_{0} & \left( P^{(0)}_{2} , E_{\mathbf{p}}^{q} \right)_{0} & \left( P^{(0)}_{3} , E_{\mathbf{p}}^{q} \right)_{0} \\
 \left( P^{(0)}_{0} , E_{\mathbf{p}}^{s} \right)_{0} & \left( P^{(0)}_{1}, E_{\mathbf{p}}^{s} \right)_{0} & \left( P^{(0)}_{2} , E_{\mathbf{p}}^{s} \right)_{0} & \left( P^{(0)}_{3} , E_{\mathbf{p}}^{s} \right)_{0}\end{array}\right)
\left( \begin{array}{c}
\Phi_{0} \\
\Phi_{1} \\
\Phi_{2} \\
\Phi_{3}  
\end{array}\right)
=
\left( \begin{array}{c}
\delta n \\
\delta \varepsilon \\
0 \\
0  
\end{array}\right).
\end{aligned}    
\end{equation}
\end{itemize}

\begin{itemize}
    \item  Massless limit: $\delta n = 0$ but $\delta \varepsilon \neq 0$ ($q=1$, $s \neq 2$) 

\begin{equation}
\label{eq:eqs-Phi's-l=0-m-0}
\begin{aligned}
\left( \begin{array}{cccc}
 \left( P^{(0)}_{0} , E_{\mathbf{p}}^{2} \right)_{0} & \left( P^{(0)}_{1}, E_{\mathbf{p}}^{2} \right)_{0} & \left( P^{(0)}_{2} , E_{\mathbf{p}}^{2} \right)_{0} \\
 \left( P^{(0)}_{0} , E_{\mathbf{p}} \right)_{0} & \left( P^{(0)}_{1}, E_{\mathbf{p}} \right)_{0} & \left( P^{(0)}_{2} , E_{\mathbf{p}} \right)_{0} \\
 \left( P^{(0)}_{0} , E_{\mathbf{p}}^{s} \right)_{0} & \left( P^{(0)}_{1}, E_{\mathbf{p}}^{s} \right)_{0} & \left( P^{(0)}_{2} , E_{\mathbf{p}}^{s} \right)_{0} \end{array}\right)
\left( \begin{array}{c}
\Phi_{0} \\
\Phi_{1} \\
\Phi_{2} \\
\end{array}\right)
=
\left( \begin{array}{c}
\delta \varepsilon \\
0 \\
0  
\end{array}\right).
\end{aligned}    
\end{equation}
\end{itemize}

\begin{itemize}
\item Massless limit: $\delta n \neq 0$ but $\delta \varepsilon = 0$ ($q =2$, $s \neq 1$)

\begin{equation}
\label{eq:eqs-Phi's-l=0-}
\begin{aligned}
\left( \begin{array}{ccccc}
 \left( P^{(0)}_{0} , E_{\mathbf{p}} \right)_{0} & \left( P^{(0)}_{1}, E_{\mathbf{p}} \right)_{0} & \left( P^{(0)}_{2}, E_{\mathbf{p}} \right)_{0} & \left( P^{(0)}_{3}, E_{\mathbf{p}} \right)_{0}  \\
 \left( P^{(0)}_{0} , E_{\mathbf{p}}^{2} \right)_{0} & \left( P^{(0)}_{1}, E_{\mathbf{p}}^{2} \right)_{0} & \left( P^{(0)}_{2} , E_{\mathbf{p}}^{2} \right)_{0} & \left( P^{(0)}_{3} , E_{\mathbf{p}}^{2} \right)_{0} \\
 \left( P^{(0)}_{0} , E_{\mathbf{p}}^{s} \right)_{0} & \left( P^{(0)}_{1}, E_{\mathbf{p}}^{s} \right)_{0} & \left( P^{(0)}_{2} , E_{\mathbf{p}}^{s} \right)_{0} & \left( P^{(0)}_{3} , E_{\mathbf{p}}^{s} \right)_{0} 
 \end{array}\right)
\left( \begin{array}{c}
\Phi_{0} \\
\Phi_{1} \\
\Phi_{2} \\
\end{array}\right)
=
\left( \begin{array}{c}
\delta n \\
0 \\
0  
\end{array}\right).
\end{aligned}    
\end{equation}
\end{itemize}
It is important to notice that in the $m \rightarrow 0$ limit, the Landau or Eckart matching conditions simply imply that $\Phi_{0} = \Phi_{1} = 0$. This is indeed what is observed in the traditional 14 moment approximation.   
Explicit solutions for the $\Phi_n$ coefficients are provided in Appendix \ref{apn:approx-moms}. Furthermore, we remark that, in the massless limit, the orthogonal polynomial basis $P^{(\ell)}_{n}$ reduces to the $n$-th order associated Laguerre polynomials $L^{(2\ell +1)}_{n}$ \cite{gradshteyn2014table}.

Now that we calculated all the expansion coefficients, for all possible matching procedures, we can express all moments of rank-0, 1, and 2 in terms of $( \delta n, \varepsilon, \delta \varepsilon, \Pi , \nu^{\mu}, h^{\mu}, \pi^{\mu \nu})$. In the massless limit, this will lead to, 
\begin{equation}
\label{eq:K-coefs-approx}
\begin{aligned}
&
\rho^{\mu \nu}_{r} \approx \mathcal{K}^{(r)}_{\pi} \beta^{-r} \pi^{\mu\nu},\\
&
\rho^{\mu}_{r} \approx  \mathcal{K}^{(r)}_{\nu} \beta^{-r} \nu^{\mu}
+
\mathcal{K}^{(r)}_{h} \beta^{1-r} h^{\mu},\\
&
\rho_{r} \approx \mathcal{K}^{(r)}_{\delta n} \beta^{1-r} \delta n 
+
\mathcal{K}^{(r)}_{\delta \varepsilon} \beta^{2-r} \delta \varepsilon.
\end{aligned}    
\end{equation}
As already mentioned, moments of rank 3 or higher simply vanish in this approximation. The expressions for the thermodynamic coefficients $\mathcal{K}^{(r)}_{\nu}, \mathcal{K}^{(r)}_{h}, \mathcal{K}^{(r)}_{\delta n},  \mathcal{K}^{(r)}_{\delta \varepsilon}$ depend on matching and will be provided in Appendix \ref{apn:approx-moms}. The coefficient $\mathcal{K}^{(r)}_{\pi}$ is matching independent and is given by,
\begin{equation}
\label{eq:K-pi-coef-approx}
\mathcal{K}^{(r)}_{\pi} 
=\beta^{r} \frac{I_{r+4,2}}{I_{4,2}}.
\end{equation}

\section{Collision term approximation}
\label{sec:5-collisionterm}

Now that we have explicitly truncated the moment expansion, we can derive the fluid-dynamical equations. In this regard, the most challenging part is simplifying the moments of the collision term, that appear in Eqs.\ \eqref{eq:transient-l=0}--\eqref{eq:transient-l=2}. For this purpose, we shall introduce another approximation, the so-called relaxation time approximation \cite{grad:1949kinetic,bhatnagar:54model,marle:69etab,welander:54temperature,andersonRTA:74}. We note that, since we are considering a derivation of the fluid-dynamical equations for arbitrary matching conditions, the traditional relaxation time approximation \cite{andersonRTA:74}, proposed by Anderson and Witting, will not suffice. This approximation does not guarantee that the conservation laws related to particle number, energy and momentum are satisfied. Instead, we follow the novel relaxation time approximation presented in \cite{rocha:21}, which is free from such flaws. This approximation reads,
\begin{equation}
\label{eq:nRTA}
\begin{aligned}
C[f_{\mathbf{p}}] \approx - \frac{E_{\mathbf{p}}}{\tau_{R}} f_{0\textbf{p}} \left[ \phi_{\mathbf{p}} - 
\frac{\left( \phi_{\mathbf{p}} , \frac{E_{\mathbf{p}}}{\tau_{R}} \right)_{0}}{(1 , \frac{E_{\mathbf{p}}}{\tau_{R}})_{0}}  
-
\frac{\left( \phi_{\mathbf{p}} , \frac{E_{\mathbf{p}}}{\tau_{R}} \tilde{P}_{1} \right)_{0}}{\left( \tilde{P}_{1} , \frac{E_{\mathbf{p}}}{\tau_{R}} \tilde{P}_{1} \right)_{0}} \tilde{P}_{1} 
- 
\frac{\left( \phi_{\mathbf{p}} , \frac{E_{\mathbf{p}}}{\tau_{R}} p^{\langle \mu \rangle} \right)_{0}}{ \left( 1 , \frac{E_{\mathbf{p}}}{\tau_{R}}\right)_{1}} p_{\langle \mu \rangle}   \right].
\end{aligned}
\end{equation}
Above, we introduced the new polynomial $\tilde{P}_{1}$ defined in such a way that it is orthogonal to $E_{\mathbf{p}}/\tau_{R}$,
\begin{equation}
\label{eq:ort2}
\left( \tilde{P}_{1} , \frac{E_{\mathbf{p}}}{\tau_{R}} \right)_{0} = 0.
\end{equation}

We further consider the following parametrization for the relaxation time, 
\begin{equation}
    \tau_{R} = t_{R} \left( \frac{E_{\mathbf{p}}}{T} \right)^{\gamma},
\end{equation}
where $t_R$ has no energy dependence. In this case, in the massless limit, $\tilde{P}^{(0)}_{1}(E_{\mathbf{p}}) = L^{(2-\gamma)}_{1}(\beta E_{\mathbf{p}}) = -\beta E_{\mathbf{p}} + (3-\gamma)$, where $L^{(\alpha)}_{n}$ denotes the $n$-th associated Laguerre polynomial. In this formulation of the relaxation time approximation, an energy-dependent $\tau_{R}$ can be used with any matching condition \cite{rocha:21} -- something that is not possible when employing the Anderson-Witting approximation for the collision term.

Using this approximation for the collision kernel, the moments of $C[f]$ can be computed in a straightforward manner. In the following, we shall calculate the relevant moments of the collision term solely in the \textit{massless limit}. In this case, the scalar moments of the collision term become
\begin{equation}
\label{eq:approx-collision-l=0}
\begin{aligned}
& \int dP E_{\mathbf{p}}^{r-1} C[f_{\mathbf{p}}] 
\approx - \frac{1}{t_{R} \beta^{\gamma}} \left\{ \rho_{r-\gamma} 
+
\frac{\Gamma(r+2-\gamma)}{\beta^{r-1}\Gamma(3-\gamma)} \left[
(r-2)\rho_{1-\gamma} 
-
\beta \frac{(r-1)}{(3-\gamma)} \rho_{2-\gamma} \right] 
\right\},
\end{aligned}
\end{equation}
the irreducible rank-1 moments of the collision term become
\begin{equation}
\label{eq:approx-collision-l=1}
\begin{aligned}
&
\int dP  E_{\mathbf{p}}^{r-1} p^{\langle \alpha \rangle} C[f_{\mathbf{p}}] 
\approx
- \frac{1}{ t_{R}\beta^{\gamma}}\left[ \rho^{\alpha}_{r-\gamma} 
-
\frac{1}{\beta^{r-1}} \frac{\Gamma(r+4-\gamma)}{\Gamma(5-\gamma)}\rho^{\alpha}_{1-\gamma}
\right] ,
\end{aligned}
\end{equation}
and the irreducible rank-2 moments of the collision term become 
\begin{equation}
\label{eq:approx-collision-l=2}
\begin{aligned}
&
\int dP  E_{\mathbf{p}}^{r-1} p^{\langle \alpha} p^{\beta \rangle} C[f_{\mathbf{p}}]  
\approx
- \frac{1}{t_{R} \beta^{\gamma}} \rho^{\alpha \beta}_{r-\gamma} .
\end{aligned}    
\end{equation}

Once the truncation of the moment expansion is applied, these collision integrals will be expressed solely in terms of $\delta n$, $\delta \varepsilon$, $\Pi$, $\nu^\mu$, $h^\mu$, and $\pi^{\mu\nu}$. The general form of these terms are 
\begin{equation}
\label{eq:collision-coefs-moms}
\begin{aligned}
\int dP  E_{\mathbf{p}}^{r-1} C[f_{\mathbf{p}}] 
& \approx
- \, \mathcal{C}_{\delta n}^{(r)} \delta n 
- \, \mathcal{C}_{\delta \varepsilon}^{(r)} \delta \varepsilon, \\
\int dP  E_{\mathbf{p}}^{r-1} p^{\langle \alpha \rangle}
C[f_{\mathbf{p}}]  
& \approx
-
\, \mathcal{C}_{\nu}^{(r)} \nu^{\alpha} 
-
\mathcal{C}_{h}^{(r)} h^{\alpha} ,
\\
\int dP  E_{\mathbf{p}}^{r-1} p^{\langle \alpha} p^{\beta \rangle}
C[f_{\mathbf{p}}] 
&
\approx
- \, \mathcal{C}_{\pi}^{(r)} \pi^{\alpha \beta} . 
\end{aligned}
\end{equation}
The transport coefficients $\mathcal{C}_{a}^{(r)}$, $ a = \delta n$, $\delta \varepsilon$, $\nu$, $h$, and $\pi$ will depend on the choice of matching conditions and the approximation imposed to the collision term. They can be found comparing Eqs.\ \eqref{eq:approx-collision-l=0}, \eqref{eq:approx-collision-l=1}, and \eqref{eq:approx-collision-l=2} with Eq.\ \eqref{eq:K-coefs-approx}, leading to
\begin{equation}
\label{eq:relations-CK}
\begin{aligned}
&
\mathcal{C}_{\delta n}^{(r)} = - \frac{1}{t_{R} \beta^{\gamma}} \left\{ \mathcal{K}^{(r-\gamma)}_{\delta n} 
+
\frac{\Gamma(r+2-\gamma)}{\beta^{r-1}\Gamma(3-\gamma)} \left[
(r-2)\mathcal{K}^{(1-\gamma)}_{\delta n} 
-
\beta \frac{(r-1)}{(3-\gamma)} \mathcal{K}^{(2-\gamma)}_{\delta n} \right] 
\right\}, \\
&
\mathcal{C}_{\delta \varepsilon}^{(r)} = - \frac{1}{t_{R} \beta^{\gamma}} \left\{ \mathcal{K}^{(r-\gamma)}_{\delta \varepsilon} 
+
\frac{\Gamma(r+2-\gamma)}{\beta^{r-1}\Gamma(3-\gamma)} \left[
(r-2)\mathcal{K}^{(1-\gamma)}_{\delta \varepsilon} 
-
\beta \frac{(r-1)}{(3-\gamma)} \mathcal{K}^{(2-\gamma)}_{\delta \varepsilon} \right] 
\right\},   \\
& \mathcal{C}_{\nu}^{(r)} = - \frac{1}{ t_{R}\beta^{r}}\left[ \mathcal{K}^{(r-\gamma)}_{\nu} 
-
\frac{1}{\beta^{r-1}} \frac{\Gamma(r+4-\gamma)}{\Gamma(5-\gamma)}\mathcal{K}^{(1-\gamma)}_{\nu} 
\right], \\ 
&
\mathcal{C}_{h}^{(r)} = - \frac{1}{ t_{R}\beta^{r}}\left[ \mathcal{K}^{(r-\gamma)}_{h} 
-
\frac{1}{\beta^{r-1}} \frac{\Gamma(r+4-\gamma)}{\Gamma(5-\gamma)}\mathcal{K}^{(1-\gamma)}_{h} 
\right],  
\\
& \mathcal{C}_{\pi}^{(r)} = - \frac{1}{t_{R} \beta^{r}} \mathcal{K}^{(r-\gamma)}_{\pi} ,
\end{aligned}    
\end{equation}
and can be explicitly calculated by inserting the truncated moment expansion for $\phi_{\mathbf{p}}$ into the collision integrals. We shall calculate explicit expressions for these transport coefficients in the next section, where we compute the relaxation times for the equations of motion.

\section{Equations of motion for the dissipative currents -- massless limit}
\label{sec:6-EoMs-m=0}

In this section we derive the equations of motion for the dissipative currents in the \textit{massless limit}. This implies that any effects due to the bulk viscous pressure will not be considered. Nevertheless, we note that, for a general matching condition this does not necessarily imply that this quantity vanishes -- it is simply replaced by $\delta \varepsilon$ as $\Pi = \delta \varepsilon/3$. The equations of motion that will be derived in the following will provide closure for the equations of motion for $n_0$, $\varepsilon_0$, and $u^\mu$ obtained from the conservation laws, see Eqs.\ \eqref{eq:hydro-EoM-n}- \eqref{eq:hydro-EoM-umu}. 

\subsection{Shear-stress tensor}
The equation of motion for the shear stress tensor is obtained by taking $r=0$ in Eq.\ \eqref{eq:transient-l=2} and substituting the results derived in Eqs.\ \eqref{eq:K-coefs-approx} and \eqref{eq:collision-coefs-moms}. One then obtains,
\begin{equation}
\label{eq:pimunu-EoM}
\begin{aligned}
& \dot{\pi}^{\left\langle \mu \nu \right\rangle}
+
\frac{\pi^{\mu \nu}}{\tau_{\pi}} 
=
\frac{8}{15} \left( \varepsilon_{0} +  \delta \varepsilon \right) \sigma^{\mu \nu} 
-
2
 \dot{u}^{\left\langle \mu \right.}  h^{\left. \nu  \right\rangle}
+
\frac{2}{5}
 \nabla^{\left\langle \mu  \right.} h^{\left. \nu \right\rangle} 
-
\frac{4}{3} \pi^{\mu \nu} \theta 
-
2 \omega_{\alpha}^{\ \left\langle \mu  \right.} \pi^{ \left. \nu  \right\rangle \alpha} 
- 
\frac{10}{7}
\sigma_{\alpha}^{\ \left\langle \mu  \right.} \pi^{ \left. \nu  \right\rangle \alpha} ,
\end{aligned}    
\end{equation}
where the shear relaxation time is identified as $\tau_\pi = 1/\mathcal{C}^{(0)}_{\pi}$. The shear relaxation time is then given by,
\begin{equation}
\label{eq:t_pi}
\frac{1}{\tau_{\pi}} = \frac{\Gamma (6-\gamma )}{120} \frac{1}{t_{R}}.
\end{equation}
If $\gamma=0$, we recover the traditional and well known result $\tau_{\pi} = t_{R}$. We note that this transport coefficient is independent of the choice of matching condition employed. 

\subsection{Particle and energy diffusion 4-currents}
The equation of motion for the particle diffusion 4-current is obtained from Eq.~\eqref{eq:transient-l=1} by taking $r=0$ and substituting the results derived in Eqs.\ \eqref{eq:K-coefs-approx} and \eqref{eq:collision-coefs-moms}. We then obtain,
\begin{equation}
\label{eq:transient-nu-mu}
\begin{aligned}
&
\dot{\nu}^{\left\langle \mu \right\rangle}
- 
\frac{\beta \dot{h}^{\left\langle \mu \right\rangle} }{4} 
+ 
\frac{\nu^{\mu}}{\tau_\nu} 
- 
\frac{\beta}{4\tau_h}h^{\mu}
=
\frac{n}{12} 
  \nabla^{\mu} \alpha 
-
\dot{u}^{\mu} \delta n  
-
\frac{1}{5}
\Delta^{\mu}_{ \ \nu} \nabla_{\alpha} \left( \beta \pi^{\nu \alpha}  \right)
+
\frac{1}{3} \nabla^{\mu}\delta n
-
\nu^{\mu} \theta
-
\omega_{\alpha}^{\ \mu} \nu^{\alpha} 
-
\frac{3}{5} \sigma_{\alpha}^{\ \mu} \nu^{\alpha}  
 \\
&
+
\frac{\beta}{4}
  \left( \Delta^{\mu}_{ \ \nu} \partial_{\alpha} \pi^{\alpha \nu}
+
h^{\alpha} \sigma_{\alpha}^{\ \mu}
+
h^{\alpha} \omega_{\alpha}^{\ \mu}
+
\frac{4}{3}
h^{\mu} \theta
-
\frac{1}{3} \nabla^{\mu} \delta \varepsilon 
+
\frac{4}{3}
\delta \varepsilon \dot{u}^{\mu} \right),
\end{aligned}
\end{equation}
where we defined the time scales $\tau_\nu = 1/\mathcal{C}^{(0)}_{\nu}$ and $\tau_h = -\beta /\left[ 4\mathcal{C}^{(0)}_{h}\right]$. These transport coefficients depend on the choice of matching condition employed. In addition, when deriving this equation we have approximated the moment $\rho_{-1}^{\mu \nu} \approx (\beta/5)\pi^{\mu \nu}$, using Eqs.\ \eqref{eq:K-coefs-approx} and \eqref{eq:K-pi-coef-approx}. It is also important to notice that, in the Landau frame, $h^{\mu} = 0$ and this equation becomes an equation of motion solely for $\nu^{\mu}$, with $\tau_{\nu}$ corresponding to a relaxation time. Meanwhile, in the Eckart frame, $\nu^{\mu} = 0$, and this equation becomes an equation of motion solely for $h^\mu$, with $\tau_{h}$ corresponding to a relaxation time. Overall, in the massless limit, we have that,  
\begin{itemize}
    \item Landau frame $(z=1)$
\begin{equation}
\frac{1}{\tau_{\nu}} = 
 \frac{\Gamma (4-\gamma )}{6} \frac{1}{t_{R}},
\end{equation}

    \item Eckart frame $(z=0)$
    \begin{equation}
    \label{eq:tR-eckart-l=1}
\frac{1}{\tau_{h}} = \frac{\Gamma (4-\gamma )}{6}  \frac{1}{t_{R}}.
    \end{equation}
\end{itemize}

We now obtain an equation of motion for the energy diffusion 4-current, $h^{\mu}$, which is required when arbitrary matching conditions are employed. Here, we extract this equation using the same moment that was employed in the matching procedure itself, see Eq.\ \eqref{eq:matching-conditions_b}. Therefore, we use Eq.\ \eqref{eq:transient-l=1} taking $r=z$ and substitute the results derived in Eqs.\ \eqref{eq:K-coefs-approx} and \eqref{eq:collision-coefs-moms}. This leads to the following equation of motion,  
\begin{equation}
\label{eq:transient-h-mu}
\begin{aligned}
&  
-
\frac{(z+3)!}{24\beta^{z-1}}
\dot{h}^{\left\langle \mu \right\rangle}
+
\mathcal{C}^{(z)}_{\nu} \nu^{\mu} + \mathcal{C}^{(z)}_{h} h^{\mu} 
=
-
\frac{n_{0}}{24\beta^{z}} (z+2)! (z-1)
  \nabla^{\mu} \alpha 
+
z \beta ^{1-z} \mathcal{K}^{(z-1)}_{\pi} 
 \pi^{\alpha \mu} \dot{u}_{\alpha}  
\\
&
-
\frac{1}{3} (z+3) \dot{u}^{\mu}  \left[ \beta^{-z}\mathcal{K}_{\delta n}^{(z+1)} \delta n
+
\beta^{1-z}\mathcal{K}_{\delta \varepsilon}^{(z+1)} \delta \varepsilon \right] 
-
\mathcal{K}^{(z-1)}_{\pi}  \Delta^{\mu}_{ \ \alpha} \nabla_{\nu} \left[ \beta ^{1-z} 
 \pi^{\nu \alpha} \right] 
+
\frac{1}{3} \nabla^{\mu}\left[  
\beta^{-z}\mathcal{K}_{\delta n}^{(z+1)} \delta n
+
\beta^{1-z}\mathcal{K}_{\delta \varepsilon}^{(z+1)} \delta \varepsilon \right]
 \\
&
+
\frac{(z+3)!}{24\beta^{z-1}} 
  \left( \Delta^{\mu}_{ \ \nu} \partial_{\alpha} \pi^{\alpha \nu}
+
h^{\alpha} \sigma_{\alpha}^{\ \mu}
+
h^{\alpha} \omega_{\alpha}^{\ \mu}
+
\frac{4}{3}
h^{\mu} \theta
-
\frac{1}{3}
\nabla^{\mu} \delta \varepsilon 
+
\frac{4}{3}
\delta \varepsilon \dot{u}^{\mu} \right).
\end{aligned}
\end{equation}
We note that in the Eckart frame, where we have $z=0$, the equation above reduces to Eq.~\eqref{eq:EoM-h-gen-eckart} as expected. The coefficients $\mathcal{C}^{(z)}_{\nu,h}$ read, 
\begin{itemize}
    \item General frame ($z \neq 0,1$)
\begin{equation}
\begin{aligned}
& \beta^{r}  t_{R} \mathcal{C}^{(r)}_{\nu} = \frac{(r-1) \beta ^{-r} \Gamma (r-\gamma +4) (-2 \gamma +r-z)}{6 z} ,
\\
&
\beta^{r-1} t_{R} \mathcal{C}^{(r)}_{h} = \frac{(r-1) \beta ^{1-r} \Gamma (r-\gamma +4) (2 \gamma -r+z-1)}{24 (z-1)} .
\end{aligned}
\end{equation}
\end{itemize}
For matching conditions in which neither diffusion 4-current is set to zero, Eqs.~\eqref{eq:transient-nu-mu} and \eqref{eq:transient-h-mu} can be diagonalized with respect to the time-like derivative terms $\dot{\nu}$ and $\dot{h}$.

\subsection{Non-equilibrium corrections to particle and energy densities}
We now derive the equations of motion satisfied by $\delta n$ and $\delta \varepsilon$. Since we are only considering the massless limit, an equation of motion for $\Pi$ is not required, since this quantity is completely determined by $\delta \varepsilon$. We note that such additional equations are not necessary for Landau or Eckart matching conditions since, in these pictures, the nonequilibrium corrections to particle and energy density simply vanish. Similarly to the procedure adopted for the rank-1 moments, we extract the equations for $\delta n$ and $\delta \varepsilon$ using the same moments that were employed in the matching procedure, see Eq.\ \eqref{eq:matching-conditions_a}. Therefore, we use Eq.\ \eqref{eq:transient-l=0} taking $r=q$ and $r=s\neq q$, leading to  
\begin{equation}
\label{eq:transient-del-n}
\begin{aligned}
& 
\frac{(q+1)!}{\beta^{q-1}} \left( \frac{q}{2} - 1\right)
 \dot{\delta n}
+
\frac{(q+1)!}{6\beta^{q-2}} (1-q) 
 \dot{\delta \varepsilon}  
 +
\frac{(q+1)!}{\beta^{q-1}} \left( \frac{q}{2} - 1\right)
 \left( \delta n \ \theta
+
\partial_{\mu} \nu^{\mu} \right) 
 \\ & 
+
\frac{(q+1)!}{6\beta^{q-2}} (1-q) \left(
u_{\mu} \dot{h}^{\mu}
+
\frac{4}{3}
\delta \varepsilon \ \theta
-
\pi^{\alpha \beta} \sigma_{\alpha \beta} 
+
\partial_{\mu} h^{\mu}
\right)
- 
q \dot{u}_{\mu} \left( \beta^{1-q} \mathcal{K}_{\nu}^{(q-1)} \nu^{\mu} 
+
\beta^{2-q} \mathcal{K}_{h}^{(q-1)} h^{\mu} \right) 
\\
&
+
\nabla_{\mu} \left( \beta^{1-q} \mathcal{K}_{\nu}^{(q-1)} \nu^{\mu} + \beta^{2-q} \mathcal{K}_{h}^{(q-1)} h^{\mu} \right)
-
(q-1) \beta^{2-q} \mathcal{K}_{\pi}^{(q-2)} \pi^{\mu \nu} \sigma_{\mu \nu} 
=
- \mathcal{C}^{(q)}_{\delta n} \delta n 
-  
\mathcal{C}^{(q)}_{\delta \varepsilon} \delta \varepsilon, 
\end{aligned}
\end{equation}
and 
\begin{equation}
\label{eq:transient-del-eps}
\begin{aligned}
&
\frac{(s+1)!}{\beta^{s-1}} \left( \frac{s}{2} - 1\right)
 \dot{\delta n}
+
\frac{(s+1)!}{6\beta^{s-2}} (1-s) 
 \dot{\delta \varepsilon}  
+
\frac{(s+1)!}{\beta^{s-1}} \left( \frac{s}{2} - 1\right)
 \left( 
\delta n \ \theta
+
\partial_{\mu} \nu^{\mu} \right) 
 \\ 
 & 
+
\frac{(s+1)!}{6\beta^{s-2}} (1-s) \left( \frac{4}{3} \delta \varepsilon \ \theta
-
\pi^{\alpha \beta} \sigma_{\alpha \beta} 
+
\partial_{\mu} h^{\mu}
+
u_{\mu} \dot{h}^{\mu}
 \right)
- 
s \dot{u}_{\mu} \left(  \beta^{1-s} \mathcal{K}_{\nu}^{(s-1)} \nu^{\mu} + \beta^{2-s} \mathcal{K}_{h}^{(s-1)} h^{\mu} \right) 
\\
&
+
\nabla_{\mu} \left( \beta^{1-s} \mathcal{K}_{\nu}^{(s-1)} \nu^{\mu} + \beta^{2-s} \mathcal{K}_{h}^{(s-1)} h^{\mu} \right)
-
(s-1) \beta^{2-s} \mathcal{K}_{\pi}^{(s-2)} \pi^{\mu \nu} \sigma_{\mu \nu} 
=
-  \mathcal{C}^{(s)}_{\delta n} \delta n 
-
\mathcal{C}^{(s)}_{\delta \varepsilon} \delta \varepsilon.  \end{aligned}
\end{equation}
Note that we obtain coupled equations of motion for $\delta n$ and $\delta \varepsilon$, even in the linear regime. When employing a matching condition that renders $\delta n = 0$, but with $\delta \varepsilon \neq 0$, (i.e. $q \neq 1,2$ and $s = 1$) Eq.\ \eqref{eq:transient-del-n} becomes a relaxation-type equation for $\delta \varepsilon$. Furthermore, for $s=1$ Eq.\ \eqref{eq:transient-del-eps} simply reduces to the continuity equation associated to the conservation of particle number, see Eq.\ \eqref{eq:hydro-EoM-n}. In this setting, we identify the following expression  for the relaxation time of $\delta \varepsilon $, 
\begin{itemize}
\item Matching Conditions with $\delta n = 0$, $\delta \varepsilon \neq 0$ ($q=1$, $s\neq 1,2$)
\begin{equation}
\label{eq:tR-q=1-l=0}
\begin{aligned}
& \frac{1}{\tau_{\delta \varepsilon}} 
\equiv
-\frac{6 \beta^{s-2}}{(s+1)!(s-1)}\mathcal{C}^{(s)}_{\delta \varepsilon}
=
\frac{\Gamma (s-\gamma +2)}{\Gamma (s+2) t_{R}}  ,
\end{aligned}    
\end{equation}
\end{itemize}
Similarly, if we consider matching conditions in which $\delta n \neq 0$, but $\delta \varepsilon = 0$, (i.e. $q \neq 1,2$ and $s = 2$) Eq.\ \eqref{eq:transient-del-n} then becomes a relaxation-type equation for $\delta n$. Furthermore, for $s=2$ Eq.\ \eqref{eq:transient-del-eps} reduces to the continuity equation associated to the conservation of energy, see Eq.\ \eqref{eq:hydro-EoM-eps}. In this case, we identify
the following expression for the relaxation time associated to $\delta n$,
\begin{itemize}
\item Matching Conditions with $\delta n \neq 0$, $\delta \varepsilon = 0$ ($q=2$, $s\neq 1,2$)
\begin{equation}
\begin{aligned}
& \frac{1}{\tau_{\delta n}} 
\equiv
\frac{\beta^{s-1}}{(s+1)!(s/2-1)}\mathcal{C}^{(s)}_{\delta n}
=
\frac{\Gamma (s-\gamma +2)}{\Gamma(s+2) t_{R}} .
\end{aligned}    
\end{equation}

For the remaining matching conditions, the collision integrals are given by,
\item General matching $\delta n \neq 0$, $\delta \varepsilon \neq 0$ ($q \neq 1,2$, $s \neq 2,1$)
\begin{equation}
\begin{aligned}
& \beta^{q-1} t_{R}  \mathcal{C}^{(q)}_{\delta n} = \frac{(q-2) (3 \gamma +s-1) \Gamma (q-\gamma +2)}{2 (s-1)}, \\
&
\beta^{q-2} t_{R} \mathcal{C}^{(q)}_{\delta \varepsilon} = 
-\frac{(q-1) (3 \gamma +s-2) \Gamma (q-\gamma +2)}{6(s-2)}.
\end{aligned}    
\end{equation}

\end{itemize}
For these more general matching conditions ($q,s \neq 1,2$) the timescales above cannot be interpreted as relaxation times for these fields. If one desires to obtain them, one must first diagonalize the equations of motion \eqref{eq:transient-del-n} and \eqref{eq:transient-del-eps}. Then, two other relaxation times arise for these coupled system of differential equations. 

\begin{figure}[!h]
    \centering
  \includegraphics[scale=0.35]{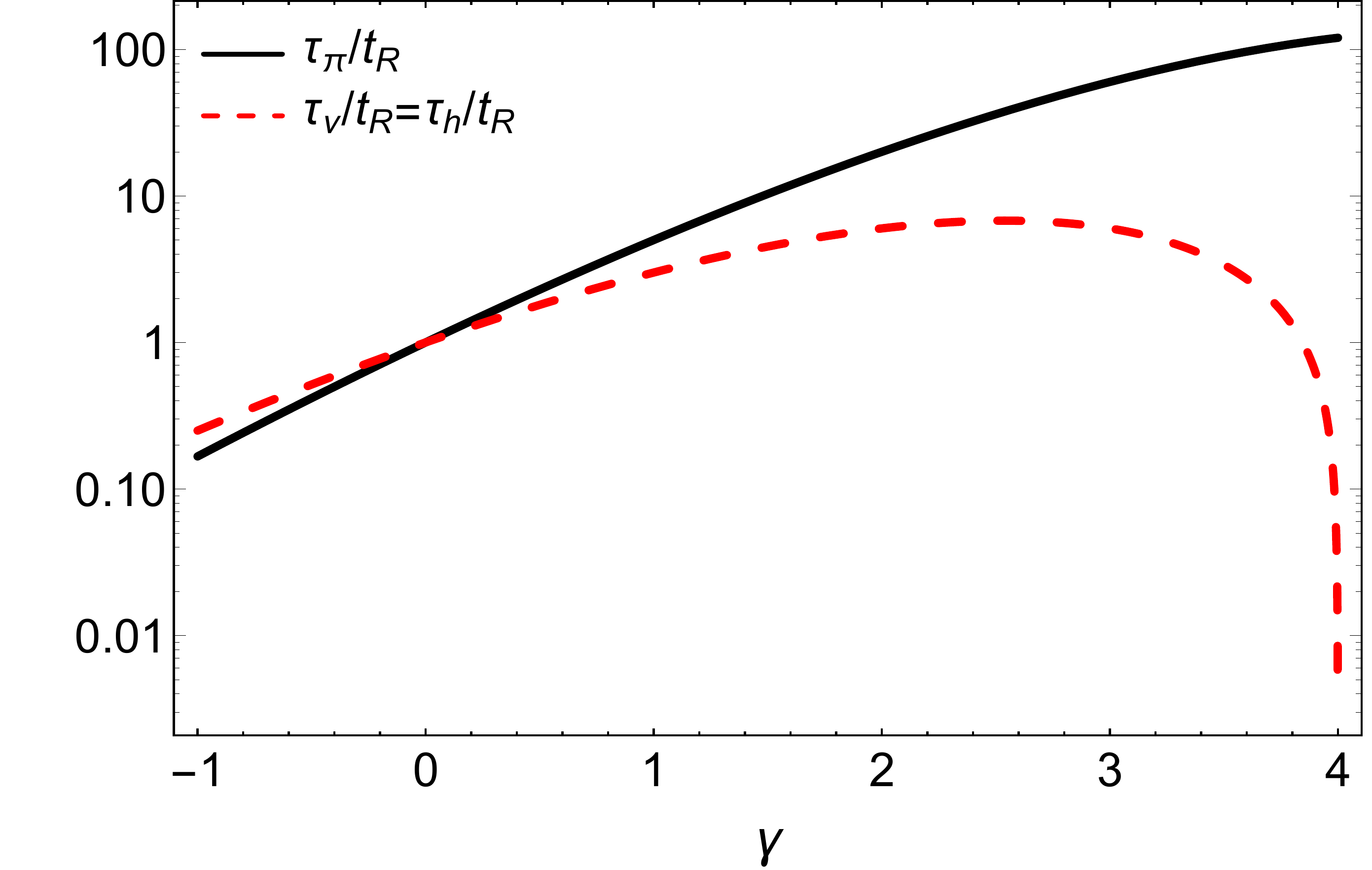}
\caption{(Color online) The relaxation times for Landau/Eckart matching conditions as a function of $\gamma$.}
\label{fig:Landau-rlx}
\end{figure}

Equations \eqref{eq:transient-nu-mu},\eqref{eq:transient-h-mu}, \eqref{eq:transient-del-n} and \eqref{eq:transient-del-eps}  reduce to the traditional Israel-Stewart equations \cite{israel1979annals} if one fixes Landau matching conditions $(q,s,z)=(1,2,1)$ \cite{denicol2012derivation}. But since we are using a more general relaxation time approximation \cite{rocha:21}, we can now calculate the transport coefficients of Israel-Stewart theory considering relaxation times that depend on energy. In Fig.\ \ref{fig:Landau-rlx} we show the relaxation times $\tau_\pi$, $\tau_\nu$ (Landau matching condition) and $\tau_h$ (Eckart matching condition), as a function of $\gamma$. We see that $\tau_{\nu}$ vanishes at $\gamma = 4$ as a result of the divergence of the integrals involving the Laguerre polynomials -- this comes about when one wants to find $x^{r-\gamma}$ as a combination of $L^{(2\ell+1)}_{n}(x)$ for $r=0$, $\gamma = 4$ and $\ell = 1$. For values above this critical one, the relaxation times become negative, evidencing the breakdown of the RTA in Eq.\ \eqref{eq:nRTA}. In its turn, for an analogous reason, $\tau_{\pi}$ vanishes at $\gamma=6$.  Finally, we note that all relaxation times become identical and equal to $t_{R}$ for $\gamma=0$.

\section{Concrete example: Exotic Eckart frame}
\label{sec:7-exotic-eckart}

For the sake of clarity, we now derive fluid-dynamical equations considering more specific matching conditions. We impose matching conditions in which the particle diffusion current is expressed as
\begin{equation}
N^{\mu} \equiv n_0 u^\mu ,
\end{equation}
as is usually the case of the Eckart frame, but the energy density does not follow the equilibrium equation of state $\delta \varepsilon \neq 0$. These matching conditions were first mentioned in Ref.\ \cite{bemfica:19nonlinear}
and we shall refer to them as Exotic Eckart frame. In practice, these set of matching conditions are implemented by taking $(q,s,z) = (1,s,0)$, $s \neq 1,2$. In this scenario, the continuity equations for $T^{\mu \nu}$ and $N^{\mu}$ become
\begin{subequations}
\begin{align}
 \dot{n}_{0} + n_{0} \theta &= 0, 
 \label{eq:n0-EoM-gen-eck}\\
 \dot{\varepsilon}_{0}+\dot{\delta \varepsilon} + \frac{4}{3} \left(\varepsilon_{0}+  \delta \varepsilon \right) \theta - \pi^{\mu \nu} \sigma_{\mu \nu} + \partial_{\mu}h^{\mu} -
 \dot{u}^{\mu} h_{\mu}  &= 0, 
  \label{eq:eps0-EoM-gen-eck}\\
\frac{4}{3} \left(\varepsilon_{0} +  \delta \varepsilon \right)\dot{u}^{\mu} -\frac{1}{3} \nabla^{\mu}\left(\varepsilon_{0} +  \delta \varepsilon \right) +  \frac{4}{3} h^{\mu} \theta + h^{\alpha}\sigma_{\alpha}^{\ \mu} + h^{\alpha}\omega_{\alpha}^{\ \mu} +\dot{h}^{\left\langle \mu \right\rangle} + \Delta^{\mu}_{\nu} \partial_{\alpha}\pi^{\alpha\nu} &= 0.
 \label{eq:umu-EoM-gen-eck}
\end{align}    
\end{subequations}
The equations of motion for the non-equilibrium correction to the energy density, $\delta \varepsilon$, the energy diffusion 4-current, $h^\mu$, and the shear-stress tensor, $\pi^{\mu \nu}$ considerably simplify and read, 
\begin{subequations}
\begin{align}
\dot{\delta \varepsilon}  
+
\frac{\delta \varepsilon }{\tau_{\delta \varepsilon}} 
&
=
-
\frac{4}{3}
\delta \varepsilon \ \theta
-
\frac{\lambda_{\delta \varepsilon \pi}}{\tau_{\delta \varepsilon}} \pi^{\alpha \beta} \sigma_{\alpha \beta}
+
\frac{\tau_{\delta \varepsilon h}}{\tau_{\delta \varepsilon}}
h^{\mu} \nabla_{\mu}P_{0}   
-
\frac{\lambda_{\delta \varepsilon h}}{\tau_{\delta \varepsilon}}
 h^{\mu} \nabla_{\mu}\alpha  
- 
\frac{\ell_{\delta \varepsilon h}}{\tau_{\delta \varepsilon}} \nabla_{\mu} h^{\mu}, 
\label{eq:EoM-del-eps-gen-eckart}
\\
 \dot{h}^{\left\langle \mu \right\rangle}
+
\frac{h^{\mu}}{\tau_h}
&
=
- 
\frac{P_{0}}{3} 
  \nabla^{\mu} \alpha 
+ 
\frac{1}{5}
\pi^{\mu \nu} 
\nabla_{\nu} \alpha 
+ 
\frac{1}{20 P_{0}}
\pi^{\mu \nu} 
\nabla_{\nu} P_{0} 
-
\frac{1}{5} \Delta^{\mu}_{ \ \nu}
\nabla_{\beta}\pi^{\nu \beta}  
-
h^{\nu} \sigma_{\nu}^{\mu}
-
h^{\nu} \omega_{\nu}^{\mu}
- 
\frac{4}{3}h^{\mu} \theta
+
\frac{1}{3} \nabla^{\mu} \delta \varepsilon 
-
\frac{1}{3}
\delta \varepsilon \nabla^{\mu}P_{0} , \label{eq:EoM-h-gen-eckart}
\\
\dot{\pi}^{\left\langle \mu \nu \right\rangle}
+ 
\frac{\pi^{\mu \nu}}{\tau_{\pi}}  
& =
\frac{8}{15} \left( \varepsilon_0 + \delta \varepsilon \right) \sigma^{\mu \nu}
-
\frac{1}{2P_{0}} h^{\left\langle \mu \right.}\nabla^{\left. \nu  \right\rangle}P_{0}
+
\frac{2}{5}
 \nabla^{\left\langle \mu  \right.} h^{\left. \nu \right\rangle} 
-
\frac{4}{3} \pi^{\mu \nu} \theta 
-
2 \omega_{\alpha}^{\ \left\langle \mu  \right.} \pi^{ \left. \nu  \right\rangle \alpha} 
- 
\frac{10}{7}
\sigma_{\alpha}^{\ \left\langle \mu  \right.} \pi^{ \left. \nu  \right\rangle \alpha} ,
\label{eq:EoM-pi-gen-eckart}
\end{align}    
\end{subequations}
where we defined the following transport coefficients,
\begin{equation}
\begin{aligned}
&
\frac{\lambda_{\delta \varepsilon \pi}}{\tau_{\delta \varepsilon}} = \frac{6}{(s+1)!}
 \mathcal{K}_{\pi}^{(s-2)}- 1,\\
& 
\frac{\lambda_{\delta \varepsilon h}}{\tau_{\delta \varepsilon}}
=
\frac{s-2}{4} \left( 1- \frac{\ell_{\delta \varepsilon h}}{\tau_{\delta \varepsilon}} \right) 
=
\frac{3}{2}\frac{(s-2)\mathcal{K}_{h}^{(s-1)}}{(s+1)!(s-1)},\\
&  \frac{\tau_{\delta \varepsilon h}}{\tau_{\delta \varepsilon}} =   
\frac{1}{2P_{0}} - \frac{\lambda_{\delta \varepsilon h}}{\tau_{\delta \varepsilon}P_{0}}\frac{2}{s-2} ,
\end{aligned}    
\end{equation}
and the remaining coefficients can be found in Eqs.~\eqref{eq:t_pi}, \eqref{eq:tR-eckart-l=1} 
and \eqref{eq:tR-q=1-l=0}. Besides, the corresponding $\mathcal{K}$-coefficients were previously introduced in Eqs.~\eqref{eq:K-coefs-approx} and \eqref{eq:K-pi-coef-approx}. When deriving Eqs.~\eqref{eq:EoM-del-eps-gen-eckart}, \eqref{eq:EoM-h-gen-eckart} and \eqref{eq:EoM-pi-gen-eckart} the equation of motion for $u^{\mu}$ was employed, 
\begin{equation}
\begin{aligned}
4 P_{0} \dot{u}^{\mu} = \nabla^{\mu}P_{0} + \cdots,
\end{aligned}    
\end{equation}
to replace the 4-acceleration by the space-like gradients of pressure. In this procedure, third-order terms in the dissipative currents and/or gradients were neglected. As it can be seen above, all the couplings can be obtained if the matching-dependent transport coefficients $\lambda_{\delta \varepsilon h}$ and $\lambda_{\delta \varepsilon \pi}$ are known. Below, we show their values for the most simple examples of Exotic Eckart matching conditions, $s=0,3,4$. 
\begin{table}[!h]
    \centering
    \begin{tabular}{|c|c|c|c|}
    \hline
      &  $s=0$ & $s=3$ & $s=4$\\
    \hline
    $\lambda_{\delta \varepsilon \pi }[\tau_{\delta \varepsilon}]$  & $- 7/10$ & $1/2$ & $11/10$\\
    \hline
    $\lambda _{\delta \varepsilon h }[\tau_{\delta \varepsilon}]$  & $-1/4$  & $5/16$ &  $3/4$ \\
    \hline     
    \end{tabular}
    \caption{Couplings in units of $\tau_{\delta \varepsilon}$ for $s=0,3,4$ Exotic Eckart matching conditions.}
    \label{tab:couplings}
\end{table}

\begin{figure}[!h]
    \centering
  \includegraphics[scale=0.35]{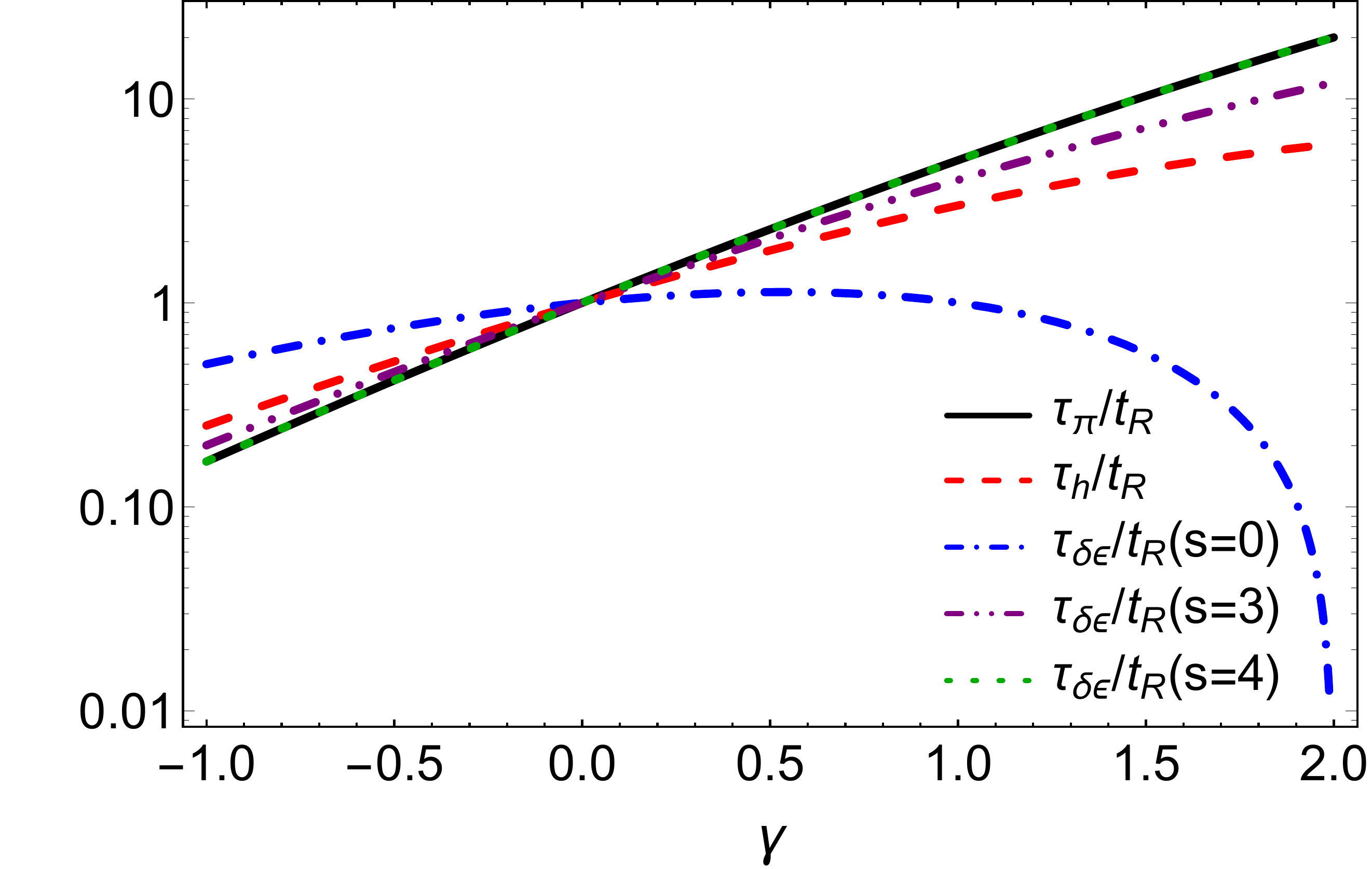}
\caption{(Color online)The relaxation times  for Exotic Eckart matching conditions.}
\label{fig:coups-and-trs}
\end{figure}

Next we plot in Fig.\ \ref{fig:coups-and-trs} the relaxation times as a function of $\gamma$ for the Exotic Eckart matching conditions for $s=0,3,4$. First, we notice that $\rho_{0}$, chosen to vanish for $s=0$, is a non-fluid-dynamic moment only in the massless limit. Otherwise, as seen in Eq.\ \eqref{eq:moments-and-hydro}, it is part of the bulk viscosity. Choosing it to vanish in general implies that non-equilibrium corrections do not change the trace of the energy-momentum tensor, $\delta T^{\mu}_{\mu} = 0$, a condition known to define the Stewart frame \cite{stewart1971non}. For $s=0$, one can see that $\tau_{\delta \varepsilon}$ vanishes at $\gamma = 2$, for the same reason already discussed in the previous section for $\tau_{h}$ and $\tau_{\nu}$. This further constrains the range of validity of the collision term ansatz of Sec.\ \ref{sec:5-collisionterm}. We note that, in general $\tau_{\delta \varepsilon}$ vanishes for larger values of $\gamma = s+2$. Once again for $\gamma = 0$ all the relaxation times reduce to $t_{R}$.  

\subsection*{Bjorken flow}
\label{sec:Bjorken}

In order to further assess the consequences of the present formulation, we solve the fluid-dynamical equations in the highly symmetric, longitudinally boost-invariant Bjorken background \cite{bjorken1983highly} assuming Exotic Eckart matching conditions. In this scenario, it is convenient to use Milne coordinates $\tau = \sqrt{t^{2}-z^{2}}$ and $\eta = \tanh^{-1}(z/t)$ so that the line element of Minkowski space reads $ds^{2} =  d\tau^{2} - dx^{2} - dy^{2} - \tau^{2} d \eta^{2}$. In these coordinates the only non-vanishing connection components are $\Gamma^{\tau}_{\eta \eta} = \tau$, $\Gamma^{\eta}_{\tau \eta}=\Gamma^{\eta}_{\eta \tau} = 1/\tau$ and a stationary fluid in these coordinates, $(u^{\tau}, u^{x}, u^{y}, u^{\eta}) = (1,0,0,0)$, translates into a longitudinally  boost-invariant expanding fluid in the usual coordinate system. The system is also assumed to be invariant under reflections around the longitudinal axis, implying that any 4-vector that is orthogonal to the fluid 4-velocity is identically zero. Therefore, the energy diffusion 4-current vanishes exactly, $h^{\mu}=0$, as well as any space-like derivative of a scalar field, e.g. $\nabla^{\mu} P_0 =\nabla^{\mu} \alpha =0$ . The coordinate system and the symmetries of the problem further imply that the shear tensor and the shear-stress tensor can be written in the following simple form, 
\begin{equation}
\begin{aligned}
&
\sigma^{\mu}_{\ \nu} =  \text{diag}\left(0, -\frac{1}{3 \tau}, -\frac{1}{3\tau}, \frac{2}{3\tau} \right),\\
&
\pi^{\mu}_{\ \nu} = \text{diag}\left(0, - \frac{\pi}{2} , - \frac{\pi}{2}, \pi \right).
\end{aligned}
\end{equation}
Finally, the expansion rate in this coordinate system is given by $\theta=1/\tau$. 

With the assumptions above, Eqs.\ \eqref{eq:eps0-EoM-gen-eck}, \eqref{eq:EoM-del-eps-gen-eckart}, and \eqref{eq:EoM-pi-gen-eckart} can be cast in the following matrix form, 
\begin{equation}
\label{eq:bjorken-eckart-matching}
\begin{aligned}
& 
\left(\begin{array}{ccc}
\dot{\varepsilon_{0}} \\
\dot{\delta \varepsilon} \\
\dot{\pi} 
\end{array}\right)
+
\left(\begin{array}{cccc}
\frac{4}{3 \tau} &  - \frac{1}{\tau_{\delta \varepsilon}}  & -\frac{1}{ \tau} \left( \frac{\lambda_{\delta \varepsilon \pi}}{\tau_{\delta \varepsilon}} + 1 \right)  \\
0 & \frac{4}{3 \tau} + \frac{1}{\tau_{\delta \varepsilon}}   & \frac{1}{\tau} \frac{\lambda_{\delta \varepsilon \pi}}{\tau_{\delta \varepsilon}}  \\
-\frac{16}{45 \tau} & -\frac{16}{45 \tau} & \frac{38}{21 \tau} + \frac{1}{\tau_{\pi}}
\end{array}\right)
\left(\begin{array}{ccc}
\varepsilon_{0} \\
\delta \varepsilon \\
\pi 
\end{array}\right)
=
\left(\begin{array}{ccc}
0 \\
0 \\
0
\end{array}\right), 
\end{aligned}
\end{equation}
where we identify that the only matching-dependent transport coefficients that remain are $\tau_{\delta \varepsilon}$ and $\lambda_{\delta \varepsilon \pi}$. We note that the equation of motion for the particle density, \eqref{eq:n0-EoM-gen-eck}, simply decouples from the equations above in Bjorken flow and will not be considered in the following \cite{Denicol:2019lio}. We solve these equations assuming that the fluid is initially in equilibrium, $\delta \varepsilon(\tau_0) = \pi(\tau_0) = 0$, with an initial time of $\tau_0=0.1$ [$t_R$]. The initial equilibrium energy density is $\varepsilon_0=1$ fm$^{-4}$. All results are shown in units of $t_R$ and, thus, they do not depend on the actual value employed for this quantity. 

\begin{figure}[!h]
    \centering
\begin{subfigure}{0.5\textwidth}
  \includegraphics[width=\linewidth]{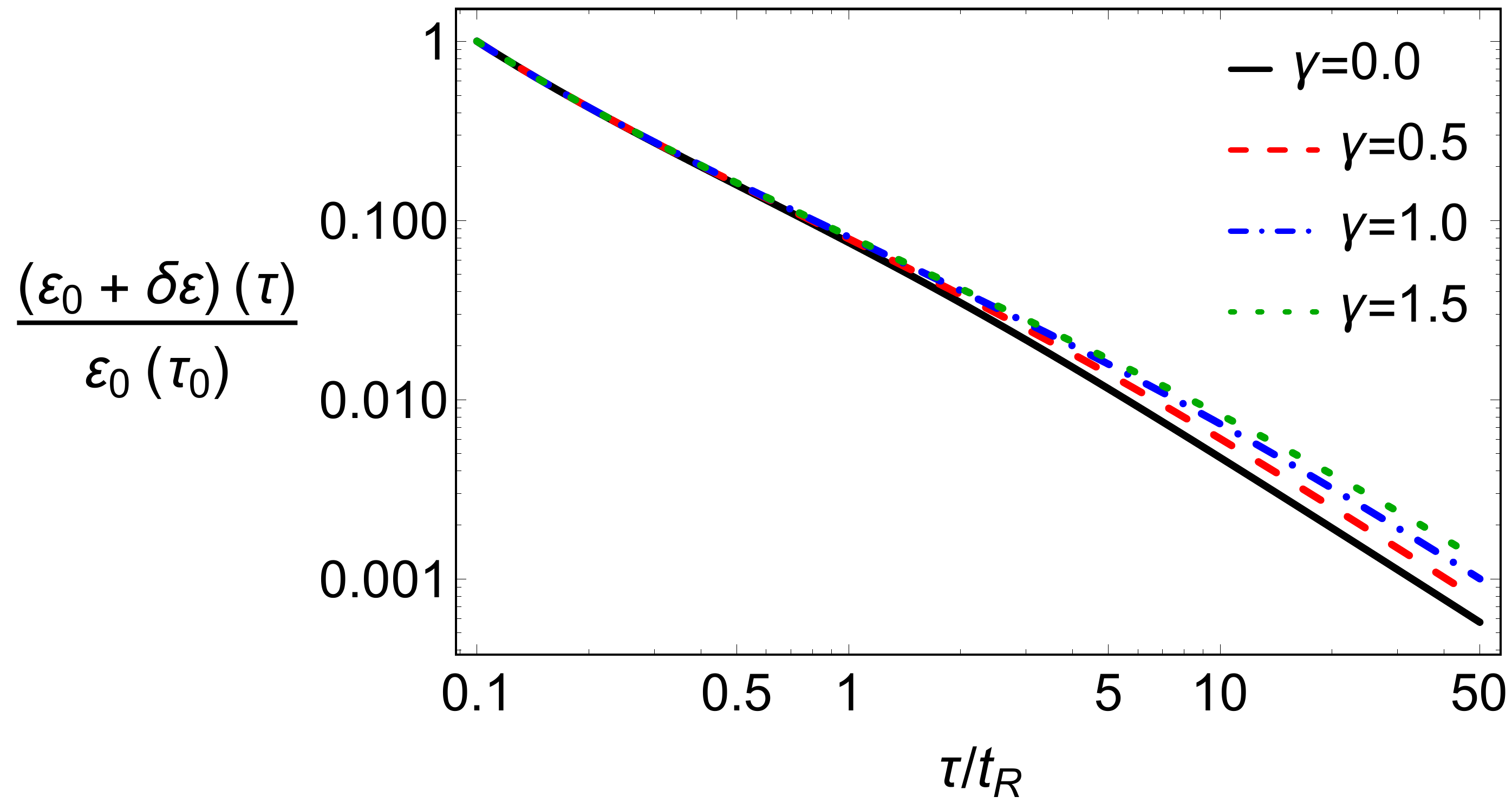}
  \caption{Total energy}
  \label{fig:tot-ener-p-eck}
\end{subfigure}\hfil 
\begin{subfigure}{0.5\textwidth}
  \includegraphics[width=\linewidth]{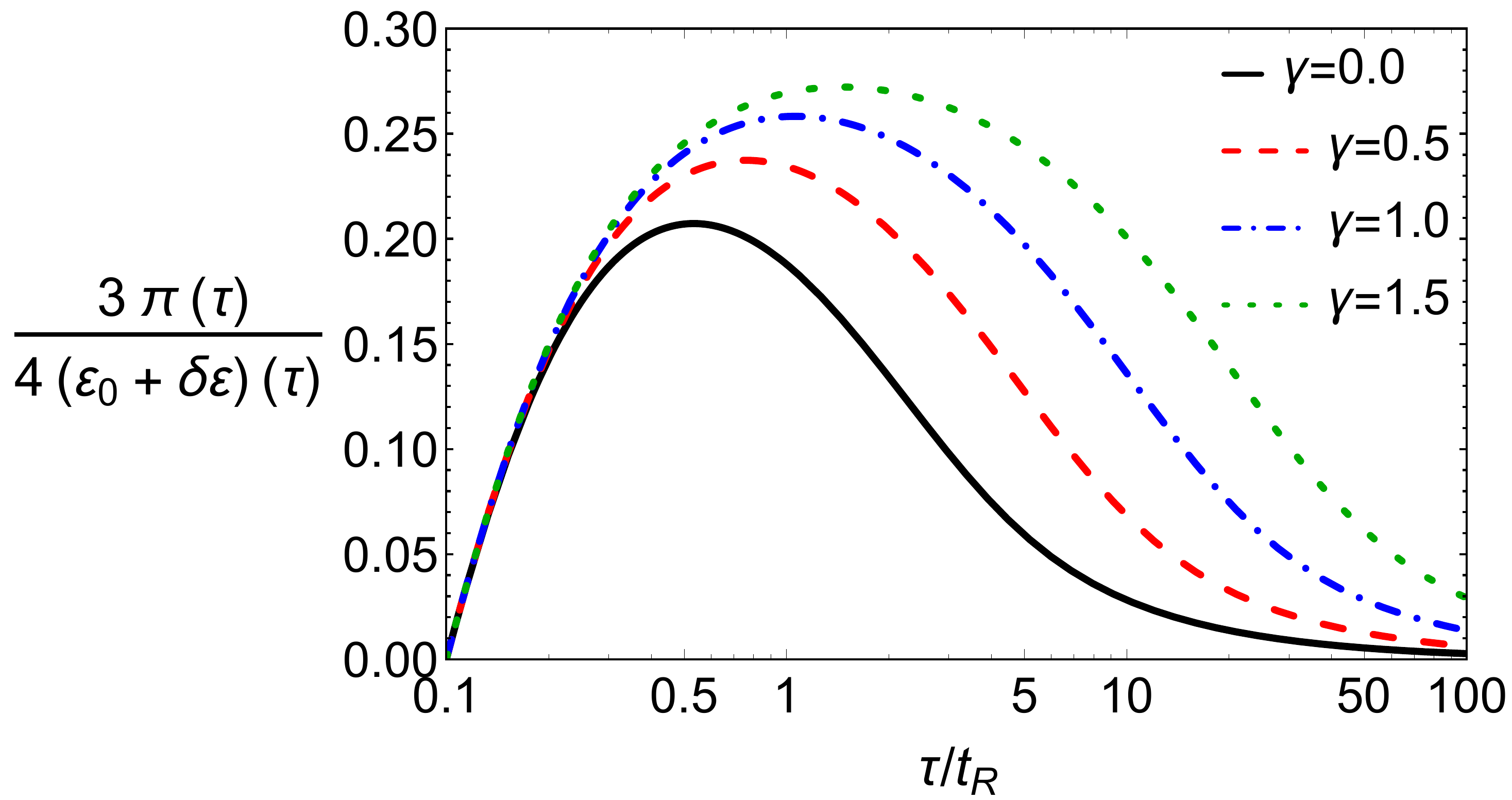}
  \caption{Anisotropic pressure}
  \label{fig:aniso-p-eck}
\end{subfigure}\hfil 
\caption{(Color online) Solutions of the transient equations for different values of $\gamma$ to the energy density and the normalized anisotropic pressure as a function of $\tau/t_{R}$ for the exotic Eckart matching conditions with $s=0$. The former is dynamically normalized by $\varepsilon + P = (4/3)(\varepsilon_{0} + \delta \varepsilon)$ }
\label{fig:transient-exotic}
\end{figure}

In Fig.~\ref{fig:transient-exotic} we show the time evolution of the total energy density and the normalized longitudinal component of the shear-stress tensor as a function of $\tau$. These results are actually matching independent since Eqs.\ \eqref{eq:hydro-EoM-eps} and \eqref{eq:pimunu-EoM} depend solely on the total energy density, $\varepsilon=\varepsilon_0 + \delta \varepsilon$, in Bjorken flow. Nevertheless, we see that the energy-dependence of the relaxation time does play a role on the time evolution of $\varepsilon$ and $\pi$. The effect of increasing $\gamma$ is to increase the timescale at which the fields evolve towards equilibrium. Increasing $\gamma$ also enhances the maximal value reached by the normalized longitudinal component of the shear-stress tensor. 
 
In Fig.~\ref{fig:T,mu-exotic-1} we plot solutions of $\delta \varepsilon$ as a function
of $\tau$, for several choices of Exotic Eckart frames, i.e., $s=0,3,4$ and different values of $\gamma$. We see that the strong longitudinal expansion leads to a significant $\delta \varepsilon$, that increases to be 20--30\% of the equilibrium energy density. Furthermore, this quantity depends significantly on the choice of matching condition, which may even lead to a change of its sign: the solution leads to positive corrections to $\varepsilon_{0}$ for $s=0$, and then negative for $s=3,4$. This change of sign can be explained by the fact that the dynamics of $\delta \varepsilon$ is dictated by its coupling with the shear-stress tensor and the transport coefficient that quantifies this coupling, $\lambda_{\delta \varepsilon \pi}$, is negative when $s=0$ and changes to positive at $s=3$ (see Table \ref{tab:couplings}). Figure \ref{fig:T,mu-exotic-1} also shows that the dynamics of $\delta \varepsilon$ can depend significantly on the energy dependence of the relaxation time, here dictated by the parameter $\gamma$. This effect can be traced to the effect of $\gamma$ on the relaxation time $\tau_{\delta \varepsilon}$, shown in Fig.\ \ref{fig:coups-and-trs}. This dependence is significantly affected by the matching condition.
\newpage
\begin{figure}[!h]
    \centering
\begin{subfigure}{0.5\textwidth}
  \includegraphics[width=\linewidth]{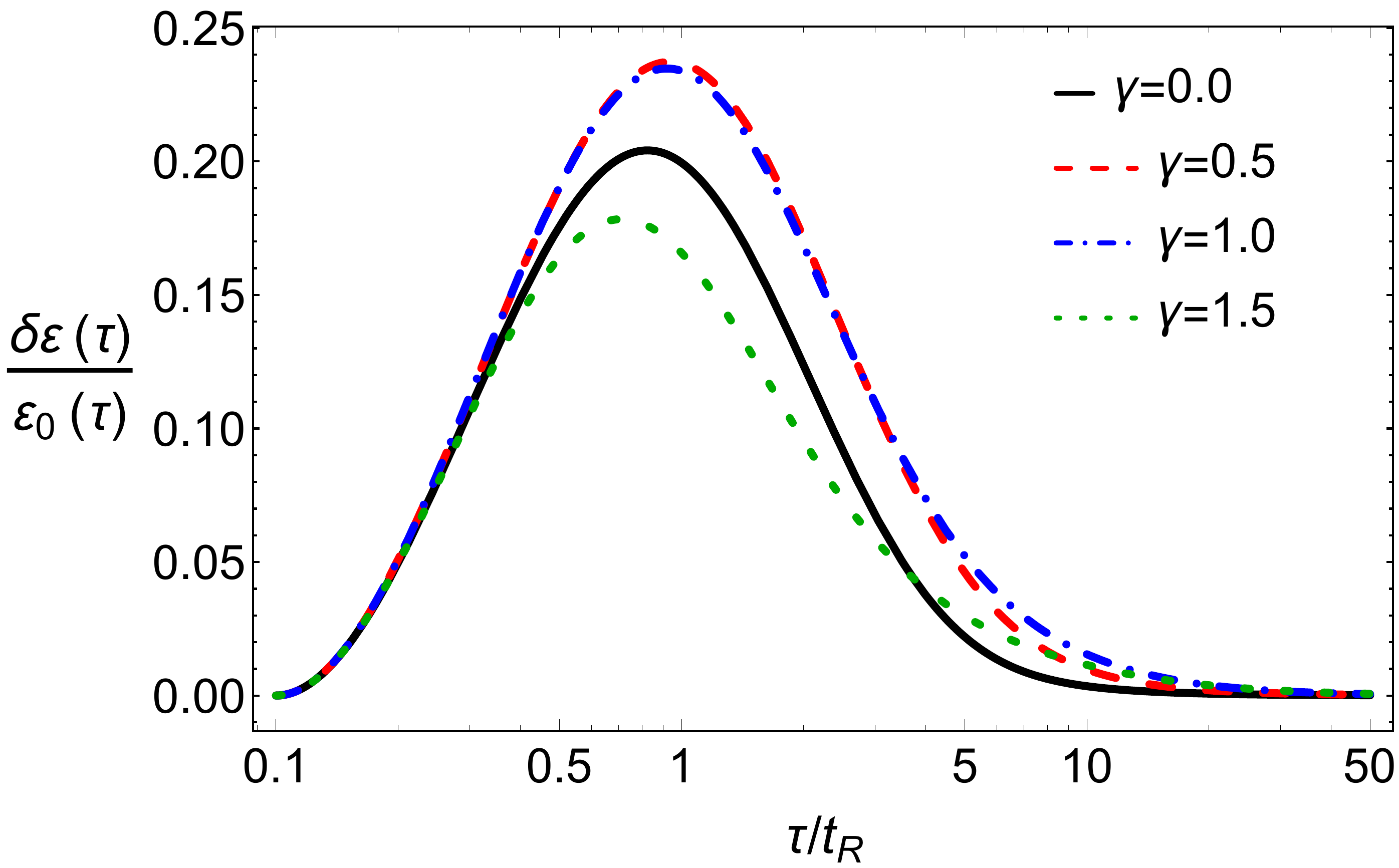}
  \caption{$s=0$}
  \label{fig:T-q=0}
\end{subfigure}\hfil
\begin{subfigure}{0.5\textwidth}
  \includegraphics[width=\linewidth]{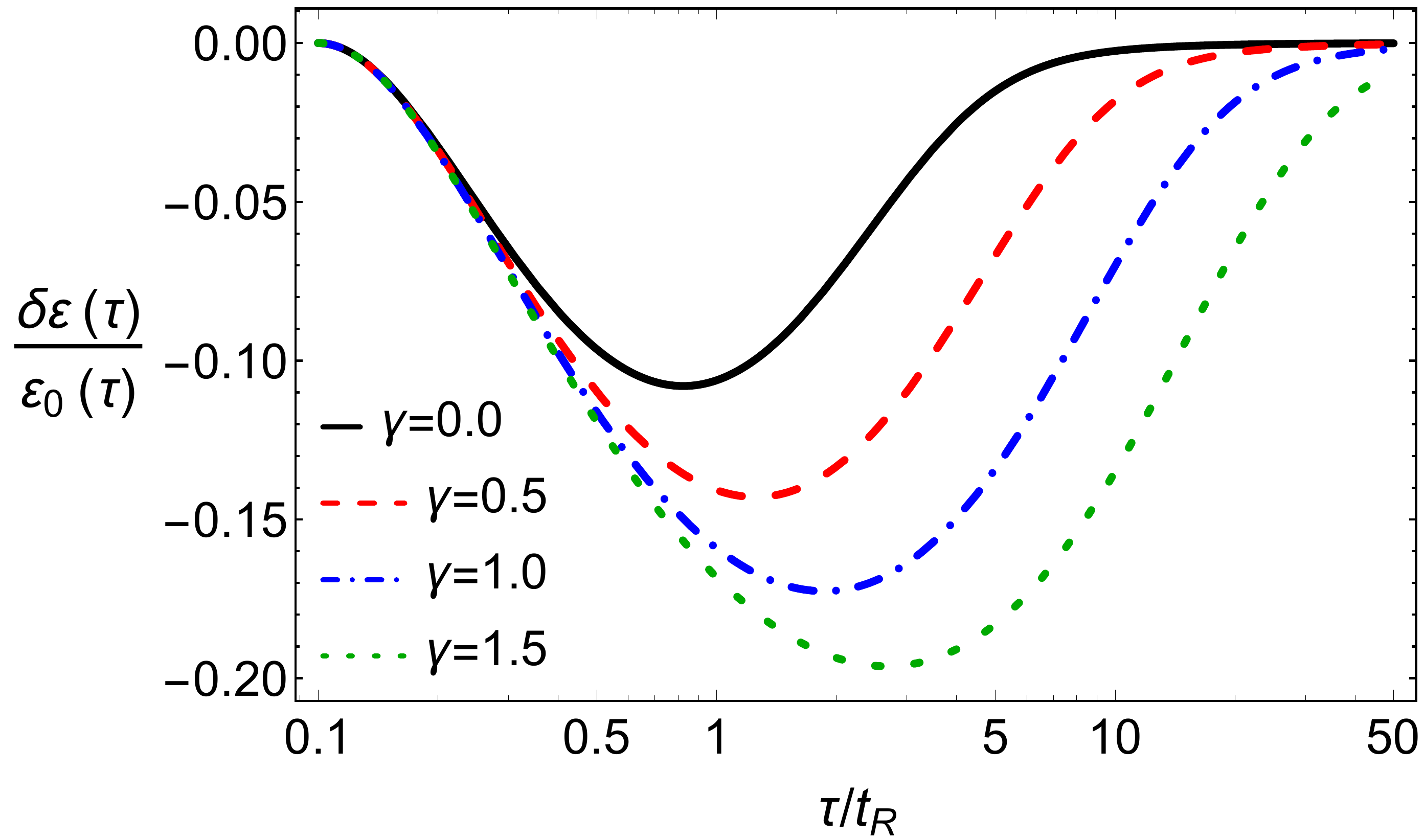}
  \caption{$s=3$}
  \label{fig:mu-q=0}
\end{subfigure}\hfil
\medskip
\begin{subfigure}{0.5\textwidth}
  \includegraphics[width=\linewidth]{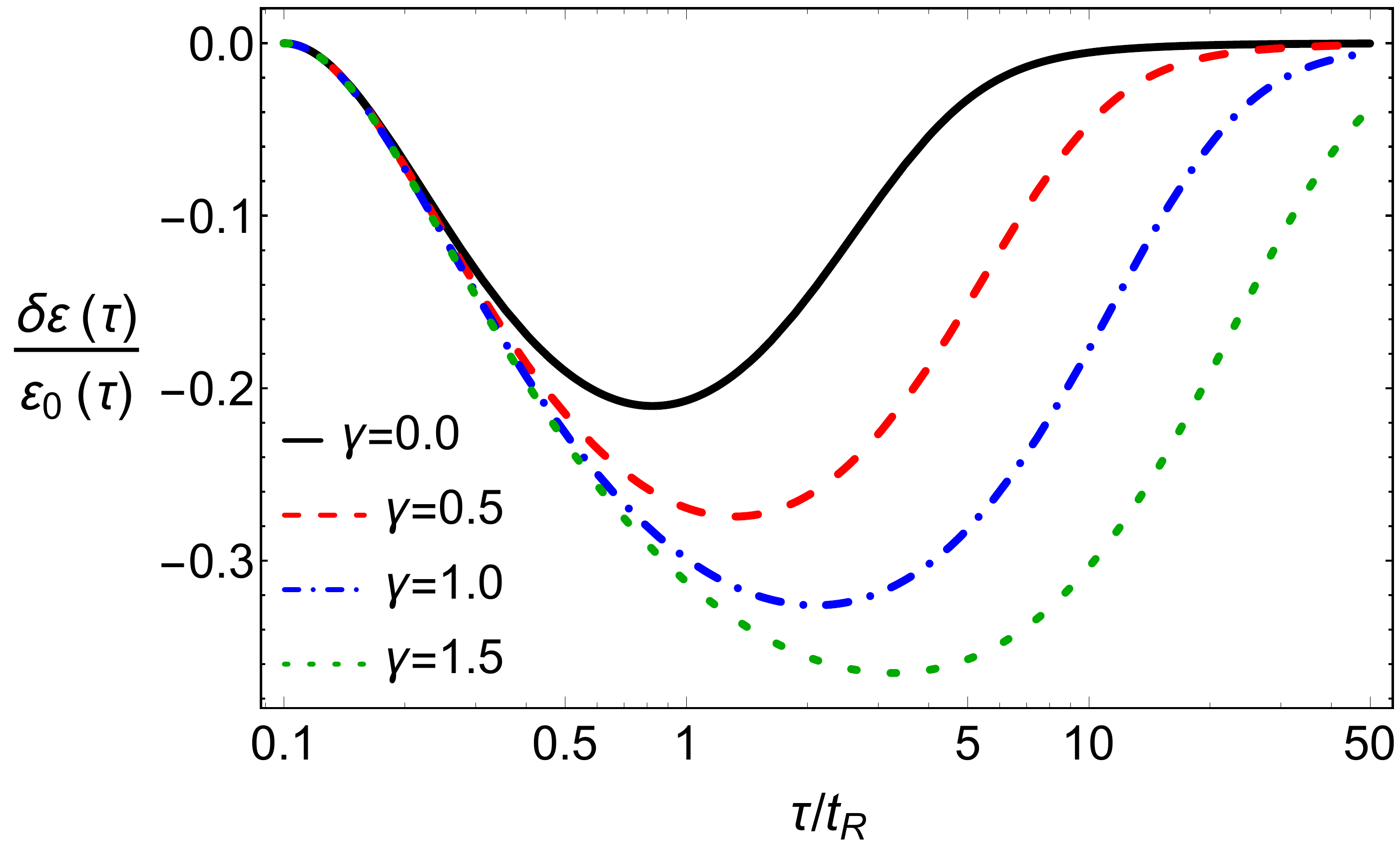}
  \caption{$s=4$}
  \label{fig:mu-q=0}
\end{subfigure}\hfil
\caption{(Color online) Solutions of the transient equations for different values of $\gamma$ to the effective temperature and chemical potential as a function of $\tau/t_{R}$ for Landau matching conditions.}
\label{fig:T,mu-exotic-1}
\end{figure}

\section{Conclusion} 
\label{sec:conclusion}

In this work we derived the Israel-Stewart equations of relativistic dissipative fluid dynamics from the Boltzmann equation assuming general matching conditions. For this purpose, we considered a single component gas of massless particles. This derivation was performed by generalizing the traditional 14-moment approximation to situations when neither the Landau nor Eckart matching conditions are imposed. 

The equations of motion derived are coupled relaxation equations for the  non-equilibrium fields $(\delta n, \delta \varepsilon, \nu^{\mu}, h^{\mu}, \pi^{\mu \nu})$, with $\delta n$ and $\delta \varepsilon$ being new dissipative quantities that may appear when general matching conditions are employed. Assuming the relaxation time approximation we calculated microscopic expressions for all the transport coefficients that appear in the fluid-dynamical equations derived in the massless limit. For the sake of illustration, solutions of these novel equations of motion were calculated using Bjorken flow and Exotic Eckart matching conditions. There we can see that the dynamics of $\delta \varepsilon$ varies appreciably with the choice of matching condition. In particular, for the rapidly expanding system described in the Bjorken flow, we see that the nonequilibrium correction to the energy density may achieve large values. 

It still remains to be verified if the equations derived in this work can be constructed to be linearly causal and stable, as happens when Landau or Eckart matching conditions are employed. Furthermore, we hope to derive these equations of motion without the assumption of massless particles and for a general collision term. These tasks will be left to future work.

\section*{Acknowledgements}

G. S. R and G. S. D. thank Jorge Noronha and Dirk Rischke for insightful discussions. G. S. R. is financially funded by Conselho Nacional de Desenvolvimento Científico e Tecnológico (CNPq), process No. 142548/2019-7. G. S. D. also acknowledges CNPq as well as Fundação Carlos Chagas Filho de Amparo à Pesquisa do Estado do Rio de Janeiro (FAPERJ), process No. E-26/202.747/2018. 

\appendix

\section{Approximation of non-fluid-dynamical and collision term  moments}
\label{apn:approx-moms}

In the moment equations \eqref{eq:transient-l=0}, \eqref{eq:transient-l=1} and \eqref{eq:transient-l=2} we see that the fluid-dynamical fields are coupled to non-fluid-dynamical moments such as the  $\rho_{-1}^{\mu}$ that appears in the equation of motion for $\nu^{\mu}$. In their turn, the collision term moments in Eqs.~\eqref{eq:approx-collision-l=0}, \eqref{eq:approx-collision-l=1} and \eqref{eq:approx-collision-l=2} involve the $\rho_{r-\gamma}^{\mu_{1} \cdots \mu_{\ell}}$ which are in general non-integer moments of the one-particle distribution function. Thus, we must be able to approximate these moments in terms of fluid-dynamical fields. This task was partially performed in Sec.\ \ref{sec:4-truncation-19} and shall be complemented in this Appendix.

We start by substituting the truncated moment expansion for $\phi_{\mathbf{p}}$, Eq.~\eqref{eq:truncation-Phi's}, into the definition of an arbitrary irreducible moment, Eq.~\eqref{eq:irreducible_moments}, leading to
\begin{equation}
\begin{aligned}
&
\rho_{r} \approx \sum_{n=0}^{N_{0}} (E_{p}^{r},P_{n}^{(0)})_{0} \ \Phi_{n}, 
\\
&
\rho^{\mu}_{r} \approx \sum_{n=0}^{N_{1}} (E_{p}^{r},P_{n}^{(1)})_{1} \ \Phi^{\mu}_{n}, 
\\
&
\rho^{\mu \nu}_{r} \approx (E_{p}^{r},1)_{2} \ \Phi^{\mu \nu}_{0}. 
\end{aligned}    
\end{equation}
The integrals $(E_{p}^{r},P_{n}^{(\ell)})_{\ell}$ can be computed analytically in the massless case, where the orthogonal polynomials reduce to associated Laguerre polynomials, $L^{(2 \ell+1)}_{n}$. In this case \cite{gradshteyn2014table},
\begin{equation}
\begin{aligned}
& (E_{p}^{r},P_{n}^{(\ell)})_{\ell} = \frac{\ell!}{(2 \ell + 1)!! } \frac{(-1)^{n + \ell}  \Gamma (r+1) \Gamma (r+2\ell+2)}{2  \beta ^{2 \ell - 2 + r} \Gamma (n+1) \Gamma (-n+r+1)} P_{0},
\end{aligned}    
\end{equation}
where we remind the reader that $P_0$ is the thermodynamic pressure, which is given by $P_{0} = e^{\alpha}/(\pi^{2} \beta^{4})$. We note that, if $r + 2 \ell + 2 = 0, -1, -2, \cdots$, the expression above diverges. 

Next, we follow the steps outlined in Sec.~\ref{sec:4-truncation-19}, one writes the expansion coefficients $\Phi_{0}^{\mu\nu}, \Phi_{0}^{\mu}, \Phi_{1}^{\mu},  \Phi_{2}^{\mu}, \Phi_{0},\Phi_{1}, \Phi_{2}, \Phi_{3}, \Phi_{4}$ as a linear combination of the fluid-dynamical fields. This procedure depends on the matching condition employed and yields the following results,
\begin{itemize}
    \item Eckart frame:
\begin{equation}
\begin{aligned}
& \Phi_{0}^{\mu} = 0, \; \; \;
\Phi_{1}^{\mu} = \frac{\beta}{4P_{0}} h^{\mu}.
\end{aligned}
\end{equation}
    \item Landau frame:
\begin{equation}
\begin{aligned}
&
\Phi_{0}^{\mu} = -\frac{1}{P_{0}} \nu^{\mu},  \; \; \;
\Phi_{1}^{\mu} = -\frac{1}{P_{0}} \nu^{\mu}. 
\end{aligned}
\end{equation}

\item General frame: 
\begin{equation}
\Phi_{0}^{\mu} = -\frac{1}{P_{0}} \nu^{\mu}, \; \; \;
\Phi_{1}^{\mu} = -\frac{1}{P_{0}} \nu^{\mu}  + \frac{\beta}{4P_{0}} h^{\mu}, \; \; \;
\Phi_{2}^{\mu} = -\frac{2}{z P_{0}} \nu^{\mu}  + \frac{\beta}{2(z-1)P_{0}} h^{\mu}.
\end{equation}
\end{itemize}

Now, for coefficients related to the scalar variables, we have,
\begin{itemize}
\item Matching conditions with $\delta n =0$, $\delta \varepsilon \neq 0$ ($q=1$ and $s \neq 1,2$) 
\begin{equation}
\begin{aligned}
& 
\Phi_{0} =
- \frac{s}{s-2} \frac{\delta \varepsilon}{\varepsilon_0}, \; \; \;
\Phi_{1} = 
 - \frac{s}{s-2} \frac{\delta \varepsilon}{\varepsilon_0}, \; \; \;
\Phi_{2} = 
-\frac{2}{s-2} \frac{\delta \varepsilon}{\varepsilon_0}.
\end{aligned} 
\end{equation}
\item Matching conditions with $\delta n \neq 0$, $\delta \varepsilon = 0$ $q = 2$ and $s \neq 2,1$
\begin{equation}
\begin{aligned}
& 
\Phi_{0} =\frac{2s}{s-1} \frac{\delta n}{n_0},
\; \; \;
\Phi_{1} = \frac{s+1}{s-1} \frac{\delta n}{n_0},
\; \; \;
\Phi_{2} = \frac{2}{s-1}  \frac{\delta n}{n_0}.
\end{aligned}    
\end{equation}
\item General frame $q,s \neq 1,2$,
\begin{equation}
\begin{aligned}
&
\Phi_{0} = 2 \frac{q s}{(q-1) (s-1)} \frac{\delta n}{n_{0}}
 -
 \frac{q s}{(q-2) (s-2)} \frac{\delta \varepsilon}{\varepsilon_{0}} \\
&
\Phi_{1} =  \frac{(q s+q+s-1)}{2 (q-1) (s-1)} \frac{\delta n}{n_{0}}
- \frac{q s}{(q-2) (s-2)} \frac{\delta \varepsilon}{\varepsilon_{0}} \\
&
\Phi_{2} = 2 \frac{(q+s-1)}{(q-1) (s-1)} \frac{\delta n}{n_{0}}
- 2 \frac{(q+s-2)}{(q-2) (s-2)} \frac{\delta \varepsilon}{\varepsilon_{0}} \\
&
\Phi_{3} = \frac{6}{(q-1) (s-1)}
\frac{\delta n}{n_{0}}
-
\frac{6}{(q-2) (s-2)} \frac{\delta \varepsilon}{\varepsilon_{0}}
\end{aligned}    
\end{equation}

\end{itemize}

Now we proceed to approximate the non-fluid-dynamical moments as
\begin{equation}
\label{eq:K-coefs-approx-}
\begin{aligned}
&
\rho^{\mu \nu}_{r} \approx \mathcal{K}^{(r)}_{\pi} \beta^{-r} \pi^{\mu\nu},\\
&
\rho^{\mu}_{r} \approx  \mathcal{K}^{(r)}_{\nu} \beta^{-r} \nu^{\mu}
+
\mathcal{K}^{(r)}_{h} \beta^{1-r} h^{\mu},\\
&
\rho_{r} \approx \mathcal{K}^{(r)}_{\delta n} \beta^{1-r} \delta n 
+
\mathcal{K}^{(r)}_{\delta \varepsilon} \beta^{2-r} \delta \varepsilon.
\end{aligned}    
\end{equation}
As the expressions for the $\Phi$ components in terms of the fluid-dynamical fields depend on matching conditions chosen, so do the $\mathcal{K}$ coefficients. Hence, for the coefficients related to the scalar fields
\begin{itemize}
    \item Generalized matching ($q=1$ $s \neq 2$)
\begin{equation}
\begin{aligned}
&  \mathcal{K}^{(z)}_{\delta \varepsilon} = \frac{(z-1) (s-z) \Gamma (z+2)}{6 (s-2)}.
\end{aligned}    
\end{equation}

    \item Generalized matching ($q=2$ $s\neq1$)
\begin{equation}
\begin{aligned}
&  \mathcal{K}^{(z)}_{\delta n} = \frac{(z-2) (z-s)  \Gamma (z+2)}{2 (s-1)}.
\end{aligned}    
\end{equation}

    \item Generalized matching ($q,s \neq 1,2$)
\begin{equation}
\begin{aligned}
&  \mathcal{K}^{(z)}_{\delta n} = \frac{(z-2) (q-z) (z-s) \Gamma (z+2)}{2 (q-1) (s-1)},
\\
&  \mathcal{K}^{(z)}_{\delta \varepsilon} = \frac{(z-1) (q-z) (s-z) \Gamma (z+2)}{6 (q-2) (s-2)}.
\end{aligned}    
\end{equation}

\end{itemize}
We should notice, again, that the coefficients above are not needed if one uses Landau/Eckart matching conditions. As for the coefficients related to the vector fields they read
\begin{itemize}
    \item Landau matching 
\begin{equation}
\label{eq:K-delta-n=0}
\begin{aligned}
&  \mathcal{K}^{(s)}_{\nu} = -\frac{1}{6} (s-1)  \Gamma (s+4). 
\end{aligned}    
\end{equation}
    \item Eckart matching
\begin{equation}
\begin{aligned}
&  \mathcal{K}^{(s)}_{h} = \frac{1}{24} s \Gamma (s+4).
\end{aligned}    
\end{equation}

    \item Generalized matching $z \neq 0,1$
\begin{equation}
\begin{aligned}
& \mathcal{K}^{(s)}_{\nu} = \frac{(s-1) (s-z) \Gamma (s+4)}{6 z}, \\
&
  \mathcal{K}^{(s)}_{h} = \frac{s (z-s) \Gamma (s+4)}{24 (z-1)}.
\end{aligned}    
\end{equation}
\end{itemize}

As for the collision term moments, $\int dP  E_{\mathbf{p}}^{r-1} p^{\langle \mu_{1}} \cdots p^{\mu_{\ell} \rangle}
C[f_{\mathbf{p}}]$, taking the prescription \eqref{eq:nRTA} into account, one must approximate the moments $\rho^{\mu_{1} \cdots \mu_{\ell}}_{r-\gamma}$, where $\gamma$ is not a integer in general. We can do so with the expressions \eqref{eq:K-coefs-approx-} with $r \mapsto r-\gamma$, then, Eqs.~\eqref{eq:relations-CK} can be derived.

\bibliographystyle{apsrev4-1}
\bibliography{liografia}

\begin{thebibliography}{39}%
\makeatletter
\providecommand \@ifxundefined [1]{%
 \@ifx{#1\undefined}
}%
\providecommand \@ifnum [1]{%
 \ifnum #1\expandafter \@firstoftwo
 \else \expandafter \@secondoftwo
 \fi
}%
\providecommand \@ifx [1]{%
 \ifx #1\expandafter \@firstoftwo
 \else \expandafter \@secondoftwo
 \fi
}%
\providecommand \natexlab [1]{#1}%
\providecommand \enquote  [1]{``#1''}%
\providecommand \bibnamefont  [1]{#1}%
\providecommand \bibfnamefont [1]{#1}%
\providecommand \citenamefont [1]{#1}%
\providecommand \href@noop [0]{\@secondoftwo}%
\providecommand \href [0]{\begingroup \@sanitize@url \@href}%
\providecommand \@href[1]{\@@startlink{#1}\@@href}%
\providecommand \@@href[1]{\endgroup#1\@@endlink}%
\providecommand \@sanitize@url [0]{\catcode `\\12\catcode `\$12\catcode
  `\&12\catcode `\#12\catcode `\^12\catcode `\_12\catcode `\%12\relax}%
\providecommand \@@startlink[1]{}%
\providecommand \@@endlink[0]{}%
\providecommand \url  [0]{\begingroup\@sanitize@url \@url }%
\providecommand \@url [1]{\endgroup\@href {#1}{\urlprefix }}%
\providecommand \urlprefix  [0]{URL }%
\providecommand \Eprint [0]{\href }%
\providecommand \doibase [0]{http://dx.doi.org/}%
\providecommand \selectlanguage [0]{\@gobble}%
\providecommand \bibinfo  [0]{\@secondoftwo}%
\providecommand \bibfield  [0]{\@secondoftwo}%
\providecommand \translation [1]{[#1]}%
\providecommand \BibitemOpen [0]{}%
\providecommand \bibitemStop [0]{}%
\providecommand \bibitemNoStop [0]{.\EOS\space}%
\providecommand \EOS [0]{\spacefactor3000\relax}%
\providecommand \BibitemShut  [1]{\csname bibitem#1\endcsname}%
\let\auto@bib@innerbib\@empty
\bibitem [{\citenamefont {Rezzolla}\ and\ \citenamefont
  {Zanotti}(2013)}]{rezzolla2013relativistic}%
  \BibitemOpen
  \bibfield  {author} {\bibinfo {author} {\bibfnamefont {L.}~\bibnamefont
  {Rezzolla}}\ and\ \bibinfo {author} {\bibfnamefont {O.}~\bibnamefont
  {Zanotti}},\ }\href@noop {} {\emph {\bibinfo {title} {Relativistic
  hydrodynamics}}}\ (\bibinfo  {publisher} {Oxford University Press},\ \bibinfo
  {year} {2013})\BibitemShut {NoStop}%
\bibitem [{\citenamefont {Takami}\ \emph {et~al.}(2015)\citenamefont {Takami},
  \citenamefont {Rezzolla},\ and\ \citenamefont {Baiotti}}]{Takami:2014tva}%
  \BibitemOpen
  \bibfield  {author} {\bibinfo {author} {\bibfnamefont {K.}~\bibnamefont
  {Takami}}, \bibinfo {author} {\bibfnamefont {L.}~\bibnamefont {Rezzolla}}, \
  and\ \bibinfo {author} {\bibfnamefont {L.}~\bibnamefont {Baiotti}},\ }\href
  {\doibase 10.1103/PhysRevD.91.064001} {\bibfield  {journal} {\bibinfo
  {journal} {Phys. Rev. D}\ }\textbf {\bibinfo {volume} {91}},\ \bibinfo
  {pages} {064001} (\bibinfo {year} {2015})},\ \Eprint
  {http://arxiv.org/abs/1412.3240} {arXiv:1412.3240 [gr-qc]} \BibitemShut
  {NoStop}%
\bibitem [{\citenamefont {Yagi}\ \emph {et~al.}(2005)\citenamefont {Yagi},
  \citenamefont {Hatsuda},\ and\ \citenamefont {Miake}}]{yagi2005quark}%
  \BibitemOpen
  \bibfield  {author} {\bibinfo {author} {\bibfnamefont {K.}~\bibnamefont
  {Yagi}}, \bibinfo {author} {\bibfnamefont {T.}~\bibnamefont {Hatsuda}}, \
  and\ \bibinfo {author} {\bibfnamefont {Y.}~\bibnamefont {Miake}},\
  }\href@noop {} {\emph {\bibinfo {title} {Quark-gluon plasma: From big bang to
  little bang}}},\ Vol.~\bibinfo {volume} {23}\ (\bibinfo  {publisher}
  {Cambridge University Press},\ \bibinfo {year} {2005})\BibitemShut {NoStop}%
\bibitem [{\citenamefont {Heinz}\ and\ \citenamefont
  {Snellings}(2013)}]{Heinz:2013th}%
  \BibitemOpen
  \bibfield  {author} {\bibinfo {author} {\bibfnamefont {U.}~\bibnamefont
  {Heinz}}\ and\ \bibinfo {author} {\bibfnamefont {R.}~\bibnamefont
  {Snellings}},\ }\href {\doibase 10.1146/annurev-nucl-102212-170540}
  {\bibfield  {journal} {\bibinfo  {journal} {Ann. Rev. Nucl. Part. Sci.}\
  }\textbf {\bibinfo {volume} {63}},\ \bibinfo {pages} {123} (\bibinfo {year}
  {2013})},\ \Eprint {http://arxiv.org/abs/1301.2826} {arXiv:1301.2826
  [nucl-th]} \BibitemShut {NoStop}%
\bibitem [{\citenamefont {Gale}\ \emph {et~al.}(2013)\citenamefont {Gale},
  \citenamefont {Jeon},\ and\ \citenamefont {Schenke}}]{Gale:2013da}%
  \BibitemOpen
  \bibfield  {author} {\bibinfo {author} {\bibfnamefont {C.}~\bibnamefont
  {Gale}}, \bibinfo {author} {\bibfnamefont {S.}~\bibnamefont {Jeon}}, \ and\
  \bibinfo {author} {\bibfnamefont {B.}~\bibnamefont {Schenke}},\ }\href
  {\doibase 10.1142/S0217751X13400113} {\bibfield  {journal} {\bibinfo
  {journal} {Int. J. Mod. Phys. A}\ }\textbf {\bibinfo {volume} {28}},\
  \bibinfo {pages} {1340011} (\bibinfo {year} {2013})},\ \Eprint
  {http://arxiv.org/abs/1301.5893} {arXiv:1301.5893 [nucl-th]} \BibitemShut
  {NoStop}%
\bibitem [{\citenamefont {Florkowski}\ \emph {et~al.}(2018)\citenamefont
  {Florkowski}, \citenamefont {Heller},\ and\ \citenamefont
  {Spalinski}}]{Florkowski:2017olj}%
  \BibitemOpen
  \bibfield  {author} {\bibinfo {author} {\bibfnamefont {W.}~\bibnamefont
  {Florkowski}}, \bibinfo {author} {\bibfnamefont {M.~P.}\ \bibnamefont
  {Heller}}, \ and\ \bibinfo {author} {\bibfnamefont {M.}~\bibnamefont
  {Spalinski}},\ }\href {\doibase 10.1088/1361-6633/aaa091} {\bibfield
  {journal} {\bibinfo  {journal} {Rept. Prog. Phys.}\ }\textbf {\bibinfo
  {volume} {81}},\ \bibinfo {pages} {046001} (\bibinfo {year} {2018})},\
  \Eprint {http://arxiv.org/abs/1707.02282} {arXiv:1707.02282 [hep-ph]}
  \BibitemShut {NoStop}%
\bibitem [{\citenamefont {Romatschke}\ and\ \citenamefont
  {Romatschke}(2019)}]{romatschke2019relativistic}%
  \BibitemOpen
  \bibfield  {author} {\bibinfo {author} {\bibfnamefont {P.}~\bibnamefont
  {Romatschke}}\ and\ \bibinfo {author} {\bibfnamefont {U.}~\bibnamefont
  {Romatschke}},\ }\href@noop {} {\emph {\bibinfo {title} {Relativistic fluid
  dynamics in and out of equilibrium: and applications to relativistic nuclear
  collisions}}}\ (\bibinfo  {publisher} {Cambridge University Press},\ \bibinfo
  {year} {2019})\BibitemShut {NoStop}%
\bibitem [{\citenamefont {Hiscock}\ and\ \citenamefont
  {Lindblom}(1985)}]{hiscock:85generic}%
  \BibitemOpen
  \bibfield  {author} {\bibinfo {author} {\bibfnamefont {W.~A.}\ \bibnamefont
  {Hiscock}}\ and\ \bibinfo {author} {\bibfnamefont {L.}~\bibnamefont
  {Lindblom}},\ }\href@noop {} {\bibfield  {journal} {\bibinfo  {journal}
  {Physical Review D}\ }\textbf {\bibinfo {volume} {31}},\ \bibinfo {pages}
  {725} (\bibinfo {year} {1985})}\BibitemShut {NoStop}%
\bibitem [{\citenamefont {Israel}\ and\ \citenamefont
  {Stewart}(1979)}]{israel1979annals}%
  \BibitemOpen
  \bibfield  {author} {\bibinfo {author} {\bibfnamefont {W.}~\bibnamefont
  {Israel}}\ and\ \bibinfo {author} {\bibfnamefont {J.}~\bibnamefont
  {Stewart}},\ }\href@noop {} {\bibfield  {journal} {\bibinfo  {journal}
  {Annals Phys}\ }\textbf {\bibinfo {volume} {118}},\ \bibinfo {pages} {228}
  (\bibinfo {year} {1979})}\BibitemShut {NoStop}%
\bibitem [{\citenamefont {Israel}(1979)}]{israel1979jm}%
  \BibitemOpen
  \bibfield  {author} {\bibinfo {author} {\bibfnamefont {W.}~\bibnamefont
  {Israel}},\ }in\ \href@noop {} {\emph {\bibinfo {booktitle} {Roy. Soc. Lond.
  A}}},\ Vol.\ \bibinfo {volume} {365}\ (\bibinfo {year} {1979})\ p.~\bibinfo
  {pages} {43}\BibitemShut {NoStop}%
\bibitem [{\citenamefont {Hiscock}\ and\ \citenamefont
  {Lindblom}(1983)}]{hiscock1983stability}%
  \BibitemOpen
  \bibfield  {author} {\bibinfo {author} {\bibfnamefont {W.~A.}\ \bibnamefont
  {Hiscock}}\ and\ \bibinfo {author} {\bibfnamefont {L.}~\bibnamefont
  {Lindblom}},\ }\href@noop {} {\bibfield  {journal} {\bibinfo  {journal}
  {Annals of Physics}\ }\textbf {\bibinfo {volume} {151}},\ \bibinfo {pages}
  {466} (\bibinfo {year} {1983})}\BibitemShut {NoStop}%
\bibitem [{\citenamefont {Denicol}\ \emph {et~al.}(2008)\citenamefont
  {Denicol}, \citenamefont {Kodama}, \citenamefont {Koide},\ and\ \citenamefont
  {Mota}}]{denicol2008stability}%
  \BibitemOpen
  \bibfield  {author} {\bibinfo {author} {\bibfnamefont {G.}~\bibnamefont
  {Denicol}}, \bibinfo {author} {\bibfnamefont {T.}~\bibnamefont {Kodama}},
  \bibinfo {author} {\bibfnamefont {T.}~\bibnamefont {Koide}}, \ and\ \bibinfo
  {author} {\bibfnamefont {P.}~\bibnamefont {Mota}},\ }\href@noop {} {\bibfield
   {journal} {\bibinfo  {journal} {Journal of Physics G: Nuclear and particle
  physics}\ }\textbf {\bibinfo {volume} {35}},\ \bibinfo {pages} {115102}
  (\bibinfo {year} {2008})}\BibitemShut {NoStop}%
\bibitem [{\citenamefont {Brito}\ and\ \citenamefont
  {Denicol}(2020)}]{brito2020linear}%
  \BibitemOpen
  \bibfield  {author} {\bibinfo {author} {\bibfnamefont {C.}~\bibnamefont
  {Brito}}\ and\ \bibinfo {author} {\bibfnamefont {G.}~\bibnamefont
  {Denicol}},\ }\href@noop {} {\bibfield  {journal} {\bibinfo  {journal}
  {Physical Review D}\ }\textbf {\bibinfo {volume} {102}},\ \bibinfo {pages}
  {116009} (\bibinfo {year} {2020})}\BibitemShut {NoStop}%
\bibitem [{\citenamefont {Biswas}\ \emph {et~al.}(2020)\citenamefont {Biswas},
  \citenamefont {Dash}, \citenamefont {Haque}, \citenamefont {Pu},\ and\
  \citenamefont {Roy}}]{biswas2020causality}%
  \BibitemOpen
  \bibfield  {author} {\bibinfo {author} {\bibfnamefont {R.}~\bibnamefont
  {Biswas}}, \bibinfo {author} {\bibfnamefont {A.}~\bibnamefont {Dash}},
  \bibinfo {author} {\bibfnamefont {N.}~\bibnamefont {Haque}}, \bibinfo
  {author} {\bibfnamefont {S.}~\bibnamefont {Pu}}, \ and\ \bibinfo {author}
  {\bibfnamefont {V.}~\bibnamefont {Roy}},\ }\href@noop {} {\bibfield
  {journal} {\bibinfo  {journal} {Journal of High Energy Physics}\ }\textbf
  {\bibinfo {volume} {2020}},\ \bibinfo {pages} {1} (\bibinfo {year}
  {2020})}\BibitemShut {NoStop}%
\bibitem [{\citenamefont {Bemfica}\ \emph {et~al.}(2018)\citenamefont
  {Bemfica}, \citenamefont {Disconzi},\ and\ \citenamefont
  {Noronha}}]{bemfica:18causality}%
  \BibitemOpen
  \bibfield  {author} {\bibinfo {author} {\bibfnamefont {F.~S.}\ \bibnamefont
  {Bemfica}}, \bibinfo {author} {\bibfnamefont {M.~M.}\ \bibnamefont
  {Disconzi}}, \ and\ \bibinfo {author} {\bibfnamefont {J.}~\bibnamefont
  {Noronha}},\ }\href@noop {} {\bibfield  {journal} {\bibinfo  {journal}
  {Physical Review D}\ }\textbf {\bibinfo {volume} {98}},\ \bibinfo {pages}
  {104064} (\bibinfo {year} {2018})}\BibitemShut {NoStop}%
\bibitem [{\citenamefont {Bemfica}\ \emph {et~al.}(2020)\citenamefont
  {Bemfica}, \citenamefont {Disconzi},\ and\ \citenamefont
  {Noronha}}]{bemfica:20general}%
  \BibitemOpen
  \bibfield  {author} {\bibinfo {author} {\bibfnamefont {F.~S.}\ \bibnamefont
  {Bemfica}}, \bibinfo {author} {\bibfnamefont {M.~M.}\ \bibnamefont
  {Disconzi}}, \ and\ \bibinfo {author} {\bibfnamefont {J.}~\bibnamefont
  {Noronha}},\ }\href@noop {} {\bibfield  {journal} {\bibinfo  {journal} {arXiv
  preprint arXiv:2009.11388}\ } (\bibinfo {year} {2020})}\BibitemShut {NoStop}%
\bibitem [{\citenamefont {Bemfica}\ \emph {et~al.}(2019)\citenamefont
  {Bemfica}, \citenamefont {Disconzi},\ and\ \citenamefont
  {Noronha}}]{bemfica:19nonlinear}%
  \BibitemOpen
  \bibfield  {author} {\bibinfo {author} {\bibfnamefont {F.~S.}\ \bibnamefont
  {Bemfica}}, \bibinfo {author} {\bibfnamefont {M.~M.}\ \bibnamefont
  {Disconzi}}, \ and\ \bibinfo {author} {\bibfnamefont {J.}~\bibnamefont
  {Noronha}},\ }\href@noop {} {\bibfield  {journal} {\bibinfo  {journal}
  {Physical Review D}\ }\textbf {\bibinfo {volume} {100}},\ \bibinfo {pages}
  {104020} (\bibinfo {year} {2019})}\BibitemShut {NoStop}%
\bibitem [{\citenamefont {Kovtun}(2019)}]{kovtun:19first}%
  \BibitemOpen
  \bibfield  {author} {\bibinfo {author} {\bibfnamefont {P.}~\bibnamefont
  {Kovtun}},\ }\href@noop {} {\bibfield  {journal} {\bibinfo  {journal}
  {Journal of High Energy Physics}\ }\textbf {\bibinfo {volume} {2019}},\
  \bibinfo {pages} {34} (\bibinfo {year} {2019})}\BibitemShut {NoStop}%
\bibitem [{\citenamefont {Hoult}\ and\ \citenamefont
  {Kovtun}(2020)}]{hoult2020stable}%
  \BibitemOpen
  \bibfield  {author} {\bibinfo {author} {\bibfnamefont {R.~E.}\ \bibnamefont
  {Hoult}}\ and\ \bibinfo {author} {\bibfnamefont {P.}~\bibnamefont {Kovtun}},\
  }\href@noop {} {\bibfield  {journal} {\bibinfo  {journal} {Journal of High
  Energy Physics}\ }\textbf {\bibinfo {volume} {2020}},\ \bibinfo {pages} {1}
  (\bibinfo {year} {2020})}\BibitemShut {NoStop}%
\bibitem [{\citenamefont {Landau}\ and\ \citenamefont
  {Lifshitz}(1959)}]{landau:59fluid}%
  \BibitemOpen
  \bibfield  {author} {\bibinfo {author} {\bibfnamefont {L.}~\bibnamefont
  {Landau}}\ and\ \bibinfo {author} {\bibfnamefont {E.}~\bibnamefont
  {Lifshitz}},\ }\href@noop {} {\bibfield  {journal} {\bibinfo  {journal}
  {Course of Theoretical Physics, Pergamon Press, London}\ }\textbf {\bibinfo
  {volume} {6}} (\bibinfo {year} {1959})}\BibitemShut {NoStop}%
\bibitem [{\citenamefont {Shen}\ \emph {et~al.}(2016)\citenamefont {Shen},
  \citenamefont {Qiu}, \citenamefont {Song}, \citenamefont {Bernhard},
  \citenamefont {Bass},\ and\ \citenamefont {Heinz}}]{Shen:2014vra}%
  \BibitemOpen
  \bibfield  {author} {\bibinfo {author} {\bibfnamefont {C.}~\bibnamefont
  {Shen}}, \bibinfo {author} {\bibfnamefont {Z.}~\bibnamefont {Qiu}}, \bibinfo
  {author} {\bibfnamefont {H.}~\bibnamefont {Song}}, \bibinfo {author}
  {\bibfnamefont {J.}~\bibnamefont {Bernhard}}, \bibinfo {author}
  {\bibfnamefont {S.}~\bibnamefont {Bass}}, \ and\ \bibinfo {author}
  {\bibfnamefont {U.}~\bibnamefont {Heinz}},\ }\href {\doibase
  10.1016/j.cpc.2015.08.039} {\bibfield  {journal} {\bibinfo  {journal}
  {Comput. Phys. Commun.}\ }\textbf {\bibinfo {volume} {199}},\ \bibinfo
  {pages} {61} (\bibinfo {year} {2016})},\ \Eprint
  {http://arxiv.org/abs/1409.8164} {arXiv:1409.8164 [nucl-th]} \BibitemShut
  {NoStop}%
\bibitem [{\citenamefont {Nunes~da Silva}\ \emph {et~al.}(2021)\citenamefont
  {Nunes~da Silva}, \citenamefont {Chinellato}, \citenamefont {Hippert},
  \citenamefont {Serenone}, \citenamefont {Takahashi}, \citenamefont {Denicol},
  \citenamefont {Luzum},\ and\ \citenamefont {Noronha}}]{NunesdaSilva:2020bfs}%
  \BibitemOpen
  \bibfield  {author} {\bibinfo {author} {\bibfnamefont {T.}~\bibnamefont
  {Nunes~da Silva}}, \bibinfo {author} {\bibfnamefont {D.}~\bibnamefont
  {Chinellato}}, \bibinfo {author} {\bibfnamefont {M.}~\bibnamefont {Hippert}},
  \bibinfo {author} {\bibfnamefont {W.}~\bibnamefont {Serenone}}, \bibinfo
  {author} {\bibfnamefont {J.}~\bibnamefont {Takahashi}}, \bibinfo {author}
  {\bibfnamefont {G.~S.}\ \bibnamefont {Denicol}}, \bibinfo {author}
  {\bibfnamefont {M.}~\bibnamefont {Luzum}}, \ and\ \bibinfo {author}
  {\bibfnamefont {J.}~\bibnamefont {Noronha}},\ }\href {\doibase
  10.1103/PhysRevC.103.054906} {\bibfield  {journal} {\bibinfo  {journal}
  {Phys. Rev. C}\ }\textbf {\bibinfo {volume} {103}},\ \bibinfo {pages}
  {054906} (\bibinfo {year} {2021})},\ \Eprint
  {http://arxiv.org/abs/2006.02324} {arXiv:2006.02324 [nucl-th]} \BibitemShut
  {NoStop}%
\bibitem [{\citenamefont {Schenke}\ \emph {et~al.}(2011)\citenamefont
  {Schenke}, \citenamefont {Jeon},\ and\ \citenamefont
  {Gale}}]{Schenke:2010rr}%
  \BibitemOpen
  \bibfield  {author} {\bibinfo {author} {\bibfnamefont {B.}~\bibnamefont
  {Schenke}}, \bibinfo {author} {\bibfnamefont {S.}~\bibnamefont {Jeon}}, \
  and\ \bibinfo {author} {\bibfnamefont {C.}~\bibnamefont {Gale}},\ }\href
  {\doibase 10.1103/PhysRevLett.106.042301} {\bibfield  {journal} {\bibinfo
  {journal} {Phys. Rev. Lett.}\ }\textbf {\bibinfo {volume} {106}},\ \bibinfo
  {pages} {042301} (\bibinfo {year} {2011})},\ \Eprint
  {http://arxiv.org/abs/1009.3244} {arXiv:1009.3244 [hep-ph]} \BibitemShut
  {NoStop}%
\bibitem [{\citenamefont {Eckart}(1940)}]{eckart:40fluid}%
  \BibitemOpen
  \bibfield  {author} {\bibinfo {author} {\bibfnamefont {C.}~\bibnamefont
  {Eckart}},\ }\href {\doibase 10.1103/PhysRev.58.919} {\bibfield  {journal}
  {\bibinfo  {journal} {Phys. Rev.}\ }\textbf {\bibinfo {volume} {58}},\
  \bibinfo {pages} {919} (\bibinfo {year} {1940})}\BibitemShut {NoStop}%
\bibitem [{\citenamefont {Chabanov}\ \emph {et~al.}(2021)\citenamefont
  {Chabanov}, \citenamefont {Rezzolla},\ and\ \citenamefont
  {Rischke}}]{chabanov:21-general}%
  \BibitemOpen
  \bibfield  {author} {\bibinfo {author} {\bibfnamefont {M.}~\bibnamefont
  {Chabanov}}, \bibinfo {author} {\bibfnamefont {L.}~\bibnamefont {Rezzolla}},
  \ and\ \bibinfo {author} {\bibfnamefont {D.~H.}\ \bibnamefont {Rischke}},\
  }\href {\doibase 10.1093/mnras/stab1384} {\bibfield  {journal} {\bibinfo
  {journal} {Monthly Notices of the Royal Astronomical Society}\ }\textbf
  {\bibinfo {volume} {505}},\ \bibinfo {pages} {5910} (\bibinfo {year}
  {2021})},\ \Eprint
  {http://arxiv.org/abs/https://academic.oup.com/mnras/article-pdf/505/4/5910/38873579/stab1384.pdf}
  {https://academic.oup.com/mnras/article-pdf/505/4/5910/38873579/stab1384.pdf}
  \BibitemShut {NoStop}%
\bibitem [{\citenamefont {Noronha}\ \emph {et~al.}(2021)\citenamefont
  {Noronha}, \citenamefont {Spali{\'n}ski},\ and\ \citenamefont
  {Speranza}}]{noronha2021transient}%
  \BibitemOpen
  \bibfield  {author} {\bibinfo {author} {\bibfnamefont {J.}~\bibnamefont
  {Noronha}}, \bibinfo {author} {\bibfnamefont {M.}~\bibnamefont
  {Spali{\'n}ski}}, \ and\ \bibinfo {author} {\bibfnamefont {E.}~\bibnamefont
  {Speranza}},\ }\href@noop {} {\bibfield  {journal} {\bibinfo  {journal}
  {arXiv e-prints}\ ,\ \bibinfo {pages} {arXiv}} (\bibinfo {year}
  {2021})}\BibitemShut {NoStop}%
\bibitem [{\citenamefont {Rocha}\ \emph {et~al.}(2021)\citenamefont {Rocha},
  \citenamefont {Denicol},\ and\ \citenamefont {Noronha}}]{rocha:21}%
  \BibitemOpen
  \bibfield  {author} {\bibinfo {author} {\bibfnamefont {G.~S.}\ \bibnamefont
  {Rocha}}, \bibinfo {author} {\bibfnamefont {G.~S.}\ \bibnamefont {Denicol}},
  \ and\ \bibinfo {author} {\bibfnamefont {J.}~\bibnamefont {Noronha}},\ }\href
  {\doibase 10.1103/PhysRevLett.127.042301} {\bibfield  {journal} {\bibinfo
  {journal} {Phys. Rev. Lett.}\ }\textbf {\bibinfo {volume} {127}},\ \bibinfo
  {pages} {042301} (\bibinfo {year} {2021})},\ \Eprint
  {http://arxiv.org/abs/2103.07489} {arXiv:2103.07489 [nucl-th]} \BibitemShut
  {NoStop}%
\bibitem [{\citenamefont {de~Groot}\ \emph {et~al.}(1980)\citenamefont
  {de~Groot}, \citenamefont {van Leeuwen},\ and\ \citenamefont {van
  Weert}}]{deGroot:80relativistic}%
  \BibitemOpen
  \bibfield  {author} {\bibinfo {author} {\bibfnamefont {S.}~\bibnamefont
  {de~Groot}}, \bibinfo {author} {\bibfnamefont {W.}~\bibnamefont {van
  Leeuwen}}, \ and\ \bibinfo {author} {\bibfnamefont {C.}~\bibnamefont {van
  Weert}},\ }\href@noop {} {\  (\bibinfo {year} {1980})}\BibitemShut {NoStop}%
\bibitem [{\citenamefont {Cercignani}\ and\ \citenamefont
  {Kremer}(2002)}]{cercignani:02relativistic}%
  \BibitemOpen
  \bibfield  {author} {\bibinfo {author} {\bibfnamefont {C.}~\bibnamefont
  {Cercignani}}\ and\ \bibinfo {author} {\bibfnamefont {G.~M.}\ \bibnamefont
  {Kremer}},\ }\href@noop {} {\emph {\bibinfo {title} {The Relativistic
  {B}oltzmann Equation: Theory and Applications}}}\ (\bibinfo  {publisher}
  {Springer},\ \bibinfo {year} {2002})\BibitemShut {NoStop}%
\bibitem [{\citenamefont {Denicol}\ \emph {et~al.}(2012)\citenamefont
  {Denicol}, \citenamefont {Niemi}, \citenamefont {Molnar},\ and\ \citenamefont
  {Rischke}}]{denicol2012derivation}%
  \BibitemOpen
  \bibfield  {author} {\bibinfo {author} {\bibfnamefont {G.}~\bibnamefont
  {Denicol}}, \bibinfo {author} {\bibfnamefont {H.}~\bibnamefont {Niemi}},
  \bibinfo {author} {\bibfnamefont {E.}~\bibnamefont {Molnar}}, \ and\ \bibinfo
  {author} {\bibfnamefont {D.}~\bibnamefont {Rischke}},\ }\href@noop {}
  {\bibfield  {journal} {\bibinfo  {journal} {Physical Review D}\ }\textbf
  {\bibinfo {volume} {85}},\ \bibinfo {pages} {114047} (\bibinfo {year}
  {2012})}\BibitemShut {NoStop}%
\bibitem [{\citenamefont {Grad}(1949)}]{grad:1949kinetic}%
  \BibitemOpen
  \bibfield  {author} {\bibinfo {author} {\bibfnamefont {H.}~\bibnamefont
  {Grad}},\ }\href@noop {} {\bibfield  {journal} {\bibinfo  {journal}
  {Communications on pure and applied mathematics}\ }\textbf {\bibinfo {volume}
  {2}},\ \bibinfo {pages} {331} (\bibinfo {year} {1949})}\BibitemShut {NoStop}%
\bibitem [{\citenamefont {Gradshteyn}\ and\ \citenamefont
  {Ryzhik}(2014)}]{gradshteyn2014table}%
  \BibitemOpen
  \bibfield  {author} {\bibinfo {author} {\bibfnamefont {I.~S.}\ \bibnamefont
  {Gradshteyn}}\ and\ \bibinfo {author} {\bibfnamefont {I.~M.}\ \bibnamefont
  {Ryzhik}},\ }\href@noop {} {\emph {\bibinfo {title} {Table of integrals,
  series, and products}}}\ (\bibinfo  {publisher} {Academic press},\ \bibinfo
  {year} {2014})\BibitemShut {NoStop}%
\bibitem [{\citenamefont {Bhatnagar}\ \emph {et~al.}(1954)\citenamefont
  {Bhatnagar}, \citenamefont {Gross},\ and\ \citenamefont
  {Krook}}]{bhatnagar:54model}%
  \BibitemOpen
  \bibfield  {author} {\bibinfo {author} {\bibfnamefont {P.~L.}\ \bibnamefont
  {Bhatnagar}}, \bibinfo {author} {\bibfnamefont {E.~P.}\ \bibnamefont
  {Gross}}, \ and\ \bibinfo {author} {\bibfnamefont {M.}~\bibnamefont
  {Krook}},\ }\href@noop {} {\bibfield  {journal} {\bibinfo  {journal}
  {Physical review}\ }\textbf {\bibinfo {volume} {94}},\ \bibinfo {pages} {511}
  (\bibinfo {year} {1954})}\BibitemShut {NoStop}%
\bibitem [{\citenamefont {Marle}(1969)}]{marle:69etab}%
  \BibitemOpen
  \bibfield  {author} {\bibinfo {author} {\bibfnamefont {C.}~\bibnamefont
  {Marle}},\ }in\ \href@noop {} {\emph {\bibinfo {booktitle} {Annales de l'IHP
  Physique th{\'e}orique}}},\ Vol.~\bibinfo {volume} {10}\ (\bibinfo {year}
  {1969})\ pp.\ \bibinfo {pages} {67--126}\BibitemShut {NoStop}%
\bibitem [{\citenamefont {Welander}(1954)}]{welander:54temperature}%
  \BibitemOpen
  \bibfield  {author} {\bibinfo {author} {\bibfnamefont {P.}~\bibnamefont
  {Welander}},\ }\href@noop {} {\bibfield  {journal} {\bibinfo  {journal}
  {Arkiv Fysik}\ }\textbf {\bibinfo {volume} {7}} (\bibinfo {year}
  {1954})}\BibitemShut {NoStop}%
\bibitem [{\citenamefont {Anderson}\ and\ \citenamefont
  {Witting}(1974)}]{andersonRTA:74}%
  \BibitemOpen
  \bibfield  {author} {\bibinfo {author} {\bibfnamefont {J.~L.}\ \bibnamefont
  {Anderson}}\ and\ \bibinfo {author} {\bibfnamefont {H.}~\bibnamefont
  {Witting}},\ }\href@noop {} {\bibfield  {journal} {\bibinfo  {journal}
  {Physica}\ }\textbf {\bibinfo {volume} {74}},\ \bibinfo {pages} {466}
  (\bibinfo {year} {1974})}\BibitemShut {NoStop}%
\bibitem [{\citenamefont {Stewart}(1971)}]{stewart1971non}%
  \BibitemOpen
  \bibfield  {author} {\bibinfo {author} {\bibfnamefont {J.~M.}\ \bibnamefont
  {Stewart}},\ }in\ \href@noop {} {\emph {\bibinfo {booktitle} {Non-equilibrium
  relativistic kinetic theory}}}\ (\bibinfo  {publisher} {Springer},\ \bibinfo
  {year} {1971})\ pp.\ \bibinfo {pages} {1--113}\BibitemShut {NoStop}%
\bibitem [{\citenamefont {Bjorken}(1983)}]{bjorken1983highly}%
  \BibitemOpen
  \bibfield  {author} {\bibinfo {author} {\bibfnamefont {J.~D.}\ \bibnamefont
  {Bjorken}},\ }\href@noop {} {\bibfield  {journal} {\bibinfo  {journal}
  {Physical review D}\ }\textbf {\bibinfo {volume} {27}},\ \bibinfo {pages}
  {140} (\bibinfo {year} {1983})}\BibitemShut {NoStop}%
\bibitem [{\citenamefont {Denicol}\ and\ \citenamefont
  {Noronha}(2020)}]{Denicol:2019lio}%
  \BibitemOpen
  \bibfield  {author} {\bibinfo {author} {\bibfnamefont {G.~S.}\ \bibnamefont
  {Denicol}}\ and\ \bibinfo {author} {\bibfnamefont {J.}~\bibnamefont
  {Noronha}},\ }\href {\doibase 10.1103/PhysRevLett.124.152301} {\bibfield
  {journal} {\bibinfo  {journal} {Phys. Rev. Lett.}\ }\textbf {\bibinfo
  {volume} {124}},\ \bibinfo {pages} {152301} (\bibinfo {year} {2020})},\
  \Eprint {http://arxiv.org/abs/1908.09957} {arXiv:1908.09957 [nucl-th]}
  \BibitemShut {NoStop}%
\end{thebibliography}%

\end{document}